\pdfoutput = 1
\documentclass[%
 reprint,
 prd,
 amsmath,amssymb,
 aps,
 superscriptaddress,
 floatfix
]{revtex4-1}

\usepackage{graphicx}
\usepackage{dcolumn}
\usepackage{bm}
\usepackage{ulem}
\usepackage{color, soul}
\graphicspath{ {./plots/} }
\usepackage{multirow}
\usepackage{placeins}

\usepackage[lofdepth,lotdepth]{subfig}

\usepackage[hidelinks]{hyperref}
\hypersetup{
	colorlinks   = true, %Colours links instead of ugly boxes
	urlcolor     = blue, %Colour for external hyperlinks
	linkcolor    = blue, %Colour of internal links
	citecolor   = red %Colour of citations
}
\newcommand\apjl{The Astrophysical Journal Letters}
\newcommand\mnras{Monthly Notices of the Royal Astronomical Society}
\newcommand\na{New Astron.}

\begin{document}

\title{The effect of data gaps on the detectability and parameter estimation 
of massive black hole binaries with LISA}
\author{Kallol Dey}
\affiliation{School of Physics, Indian Institute of Science Education and Research Thiruvananthapuram, Maruthamala PO, Vithura, Thiruvananthapuram 695551, Kerala, India}
\author{Nikolaos Karnesis}
\affiliation{Department of Physics, Aristotle University of Thessaloniki, Thessaloniki 54124, Greece}
\affiliation{APC, AstroParticule et Cosmologie, 
Université de Paris, CNRS, Astroparticule et Cosmologie, F-75013 Paris, France}
\author{Alexandre Toubiana}
\affiliation{APC, AstroParticule et Cosmologie, 
Université de Paris, CNRS, Astroparticule et Cosmologie, F-75013 Paris, France}
\author{Enrico Barausse}
\affiliation{SISSA, Via Bonomea 265, 34136 Trieste, Italy and INFN Sezione di Trieste, 34127 Trieste, Italy}
\affiliation{IFPU - Institute for Fundamental Physics of the Universe, Via Beirut 2, 34014 Trieste, Italy}
\author{Natalia Korsakova}
\affiliation{ARTEMIS, Observatoire de la Côte d’Azur, Bd de l'Observatoire, BP 4229, 06304 Nice, France}
\affiliation{SYRTE, Observatoire de Paris, Université PSL, CNRS, Sorbonne Université, LNE, 75014 Paris, France}
\author{Quentin Baghi}
\affiliation{CEA, Centre de Saclay, IRFU/DPhP, 91191 Gif-sur-Yvette, France}
\author{Soumen Basak}
\affiliation{School of Physics, Indian Institute of Science Education and Research Thiruvananthapuram, Maruthamala PO, Vithura, Thiruvananthapuram 695551, Kerala, India}

%\date{\today}

\begin{abstract}
Massive black hole binaries are expected to provide the strongest gravitational wave signals for the Laser Interferometer Space Antenna (LISA), a space mission targeting $\sim\,$mHz frequencies.
As a result of the technological challenges inherent in the mission's design, implementation and long duration (4 yr nominal), the LISA data stream is expected to be affected 
by relatively long  gaps where no data is collected (either because of hardware failures, or 
because of scheduled maintenance operations, such as re-pointing of the antennas toward the Earth). 
Depending on their mass,
massive black hole binary signals
may range from
quasi-transient to very long lived, and it is unclear how data gaps will impact detection and parameter estimation of these sources. Here, we will explore this question by using state-of-the-art astrophysical models for the population of massive black hole binaries. We will investigate the potential detectability of MBHB signals by observing the effect of gaps on their signal-to-noise ratios. We will also assess the effect of the gaps on parameter estimation for these sources, using the Fisher Information Matrix formalism as well as full Bayesian analyses. Overall, we find that the effect of data gaps due to regular maintenance of the spacecraft is negligible, except for systems that coalesce within such a gap. The effect of unscheduled gaps, however, will probably be more significant than that of scheduled ones.
\end{abstract}

\maketitle

\setlength{\columnsep}{0.5cm}

\section{Introduction}

The detection of gravitational waves (GW) by the LIGO-Virgo collaboration \cite{PhysRevLett.116.061102, PhysRevLett.116.241103, PhysRevX.6.041015, PhysRevLett.118.221101, 2017ApJ...851L..35A, PhysRevLett.119.141101, PhysRevLett.119.161101, 2020ApJ...892L...3A, PhysRevD.102.043015, PhysRevLett.125.101102} has been hailed as one of the most significant scientific advancements of our time, and rightly so as it opened up a new way of probing the cosmos - using GWs in addition to electromagnetic waves. LIGO, Virgo and other similar ground based observatories are sensitive to GWs in the frequency range from 15 Hz to few kHz \cite{PhysRevX.9.031040,2018LRR....21....3A},
while pulsar-timing array (PTA) \cite{Foster:apj:1990} experiments like the European Pulsar Timing Array (EPTA) \cite{Kramer_2013}, the Parkes Pulsar Timing Array (PPTA) \cite{Hobbs_2013}, the North American  Nanohertz Observatory  for Gravitational Waves (NANOGrav) \cite{McLaughlin_2013} and the International Pulsar Timing Array (IPTA) \cite{Hobbs_2010, Manchester_2013} probe frequencies ranging from nHz to $\mu$Hz. To detect frequencies intermediate between these two extremes, one will have to resort to space-based GW detectors.
The Laser Interferometer Space Antenna (LISA)~\cite{2017arXiv170200786A} is such a detector, which will be sensitive to GW radiation from various types of sources in the range 0.1 mHz to 1 Hz. Among those sources are  massive black hole binaries (MBHBs) \cite{PhysRevD.93.024003,Bonetti:2018tpf,2020ApJ...904...16B, Sesana_2005,10.1111/j.1365-2966.2007.11734.x, Sesana_2009, Arun_2009}, which are the main focus of this paper. One of the key differences between LISA and LIGO/Virgo is that the sources that LISA will be sensitive to will remain in its bandwidth for long times, ranging from hours to years. The observatory itself will be operational for four years of nominal duration, and possibly longer. The
technological challenges inherent in the mission and also its 
long  duration 
mean that it is important to consider the possibility that parts of the spacecraft might face unexpected problems. In addition, LISA  will also need to undergo regular maintenance. All this will result in periods of time when no useful data will be collected. We refer to these periods as ``data gaps''. Owing to the long lifetime of the signals in the LISA bandwidth,
these gaps will not only affect portions of a given signal, 
but they may pose serious problems in the analysis of the recorded data. 

In order to test the  technological developments needed to meet the requirements of LISA, the European Space Agency (ESA) launched LISA Pathfinder (LPF) in December 2015 \cite{2019arXiv190308924A}. It was operational from February 2016 to June 2017. The observations performed by LPF highlight 
the need to study the impact of data gaps on  GW signals. LPF results suggest that there may indeed be  disruptions in the data stream
in the course of the mission. In addition, there will also be scheduled gaps to ensure regular maintenance of the on-board apparatus. The foreseen scenario for the scheduled gaps may be, for example, about 3.5 hours every week, or about 7 hours every 2 weeks.

Previous studies on data gaps in LISA signal have been restricted to simple models for gaps (and GW signals) \cite{Pollack2004, CarrePorter2010}, with realistic gaps being studied only recently with the knowledge of LPF measurements \cite{PhysRevD.100.022003}. Even now, MBHB signals have largely remained untouched in this regard.
In this paper, we will explore how the presence of gaps (both scheduled and unscheduled) in the data stream will affect our analyses of MBHB signals. 
We utilize state-of-the-art predictions for the astrophysical population of massive black hole binaries (MBHBs) detectable by LISA, and
study the impact of gaps on the analysis of GW signals originating from these sources. Specifically, we study the impact of gaps on the (potential) detectability and parameter estimation of synthetic sources produced with eight different models of MBHB populations (see Sec.~\ref{astro_catalogs}). For detectabilty, we look at the effect of the gaps on 
the signal-to-noise ratio (SNR), while for parameter estimation
we use the
Fisher Information Matrix (FIM) formalism, as well as full Bayesian analyses for
specific MBHBs. It should be noted that the  SNR is only one of the many factors that impact the detection of a GW signal, especially for LISA (for which several overlapping sources will be present in the data). In this paper, however, we will optimistically consider the SNR as the only metric for the detectability of a signal. 
Similarly, for parameter estimation we analyse each signal individually, neglecting any other possible overlapping signals.
Overall, we find that the 
long duration of MBHB signals typically leads to the gaps having little to no bearing, at least as long as the merger does not take place exactly within a gap. 
We also find, however, that unscheduled gaps have a more prominent effect than 
scheduled ones.

This paper is organized as follows. Sec.~\ref{types_of_gaps} introduces the different types of gaps that LISA is likely to experience and their possible causes. Sec.~\ref{data_prepare} describes how gaps are simulated for this study. In Sec.~\ref{reponse_and_noise} we discuss, in brief, the LISA response function, the noise and waveform models considered. Sec.~\ref{methodology} presents our methods for studying the impact of gaps on the analysis of GW signal. Sec.~\ref{astro_catalogs} introduces the various astrophysical catalogs of MBHBs under consideration. In Sec.~\ref{impact_of_gaps} we present our results for the detectability and parameter estimation of MBHBs in the presence of gaps running a Fisher matrix analysis on the astrophysical catalogs, and in Sec.~\ref{bayesian_PE} we focus on two systems for which we perform a full Bayesian analysis with and without gaps. We summarize our findings in Sec.~\ref{conclusion}.

\section{Types of gaps in data}\label{types_of_gaps}

Scheduled gaps correspond to the disruptions due to routine maintenance of the LISA spacecraft and the instruments on board. For instance, the on-board antenna, which allows the instruments to contact Earth, needs to be re-oriented periodically to maintain the communication with the spacecraft. The rotation of the antenna will introduce additional noise which will make data unusable for scientific analysis. These disturbances will lead to a small gap of 3.5 hours in the data stream every week. Alternatively, we could also choose to perform maintenance operations on a biweekly basis, resulting in a 7 hour long gap every two weeks.  In addition to these maintenance induced gaps, there may also be other missing data due to unforeseen anomalies and/or failures in one or more components in the instrument. As the exact nature of these anomalies and faults is unknown, so is the time needed to bring the system back to science mode. It is hypothesized that most of the issues will take at least a day to fix, with some taking as long as several days. These type of breaks in usable data will be referred to as unscheduled gaps.

Unscheduled random gaps can happen due to hardware problems or unexpected physical events. The observations made by LPF provide some useful insight in this regard. We describe the procedure of how we estimated the length and duration of unscheduled gaps in Appendix~\ref{app:dfacs}.
As we shall see in the next sections, unscheduled gaps appear to have greater impact on the signal measurements than scheduled gaps. The observations made by LISA Pathfinder (LPF) indicate that it may be feasible to require the three satellites that make up the LISA constellation to simultaneously collect data for 75 \% of the mission duration.

Although the exact duration of the gaps in the data and the intervals between them are unknown, reasonable estimates can be put forward that comply with the demand for 75\% duty cycle. In this paper, we simulate the presence of unscheduled gaps by considering individual gaps, each of which is 3 days long, with the intervals between successive gaps drawn from an exponential distribution whose scale factor is determined by the duty cycle 75\% requirement. The duration is set under the assumption that problems that take more than 3 days to solve will be sparse during the LISA mission. This, however, is just an estimate and the true nature of the unscheduled gaps will not be known until LISA is fully operational. It is expected that our ability to solve issues will improve as the mission progresses and the time needed to resume operations after encountering a problem will also decrease accordingly. The first year is likely to have a lower duty cycle compared to the later years. Unless new and unseen anomalies keep appearing, prior experiences should allow us to manage these disruptions more efficiently. 

\section{Preparing the data}\label{data_prepare}

\begin{figure*}
\subfloat[Full distribution of Scheduled gaps \label{fig:win_sche_full}]{%
  \centering
  \includegraphics[width=0.49\textwidth,height=3.5cm]{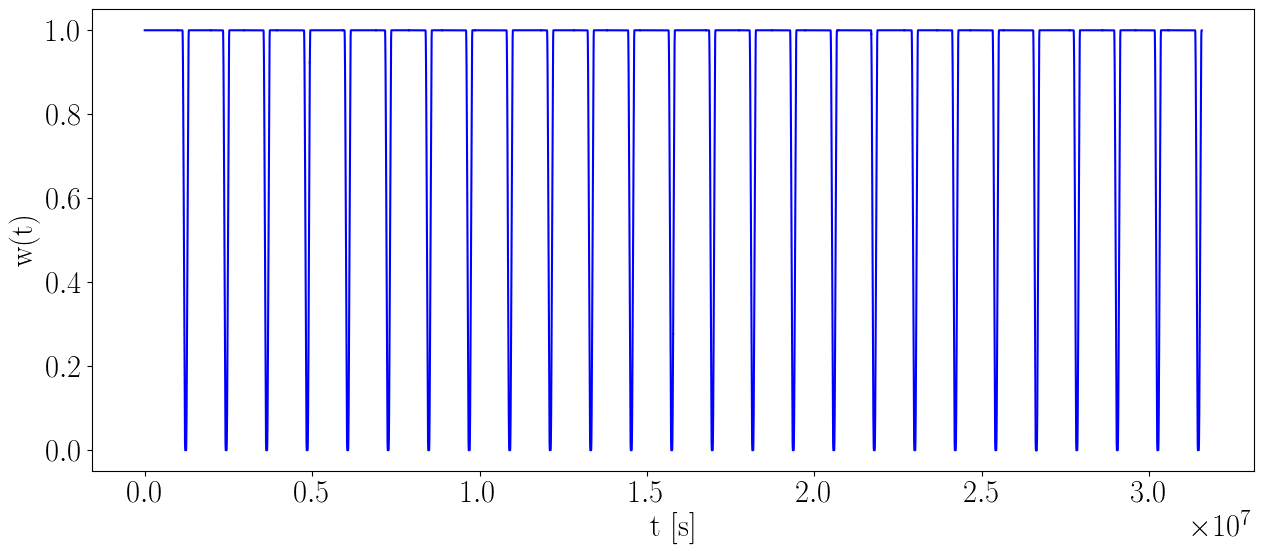}%
}\hfill
\subfloat[Individual scheduled gap \label{fig:win_sche_zoom}]{%
  \centering
  \includegraphics[width=0.49\textwidth,height=3.5cm]{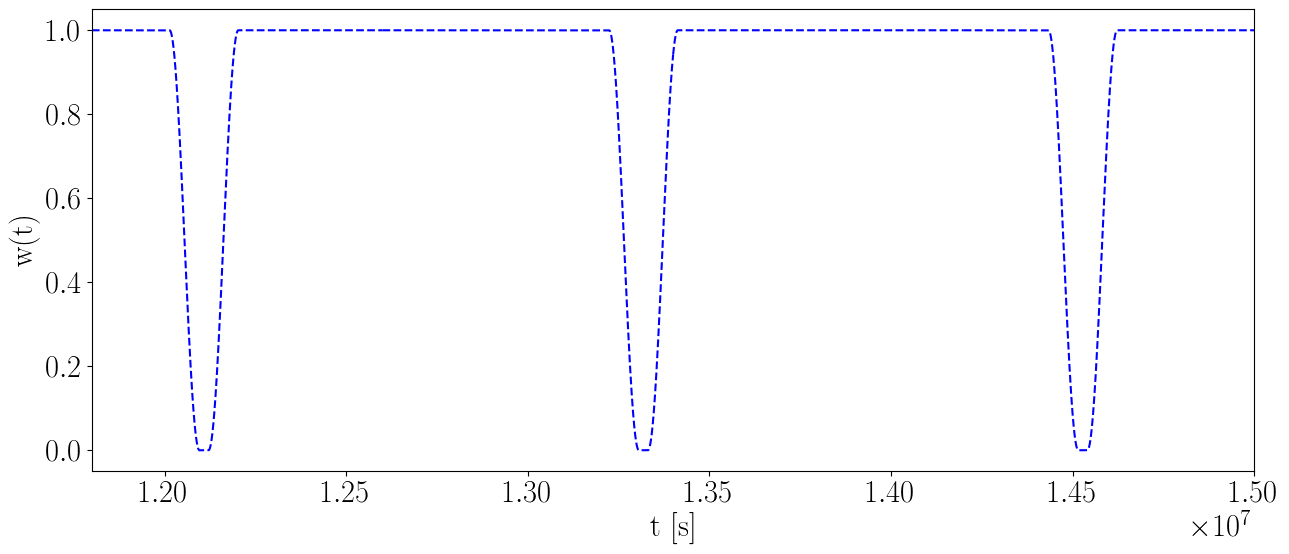}%
}\hfill
\subfloat[Full distribution of Unscheduled gaps \label{fig:win_unsche_full}]{%
  \centering
  \includegraphics[width=0.49\textwidth,height=3.5cm]{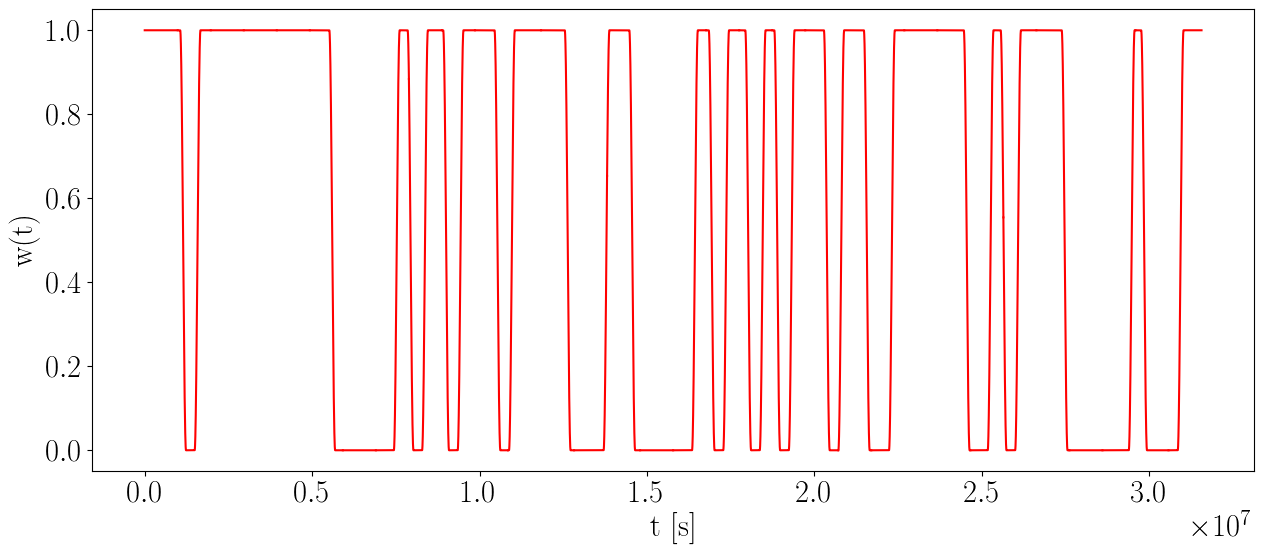}%
}\hfill
\subfloat[Individual unscheduled gap \label{fig:win_unsche_zoom}]{%
  \centering
  \includegraphics[width=0.49\textwidth,height=3.5cm]{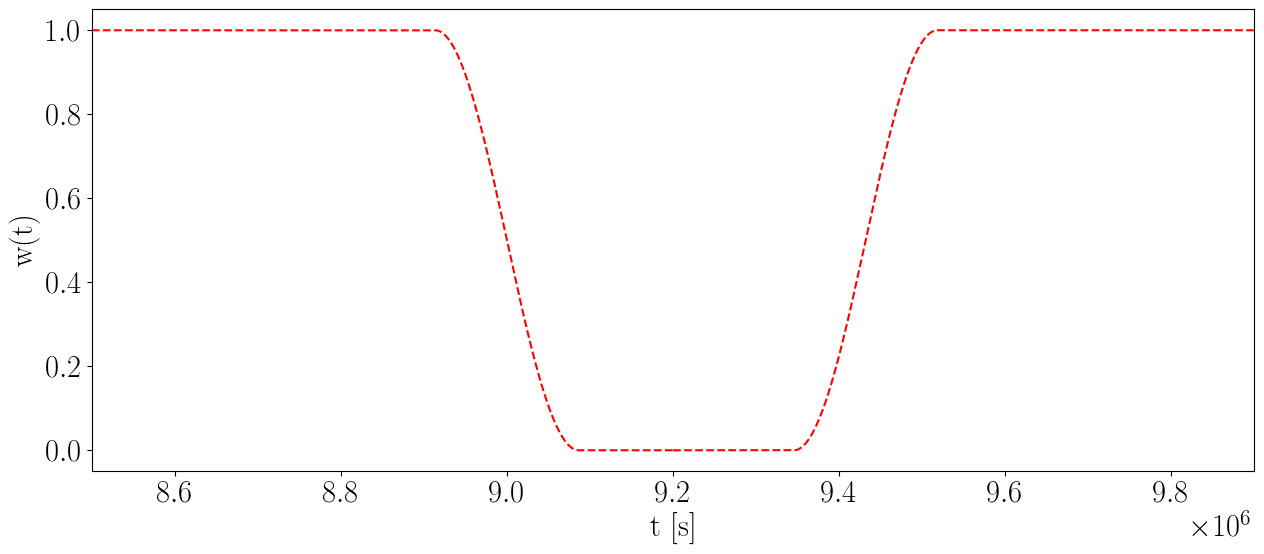}%
}\hfill
\caption{(a) and (c) show a realization of the full distribution of gaps for the scheduled and unscheduled gaps respectively. Some of the individual gaps were combined to facilitate better smoothing in case of unscheduled gaps. The individual gaps are show in (b) and (d)}
\label{fig:smooth_windows}
\end{figure*}

We begin by simulating the effect of gaps in the data stream. This is done by \textit{windowing} the GW signal with a suitable function that describes the gaps. We should mention that this treatment, although simple, leads to greater loss in data, as is explained later in this section. This does, however, form a good baseline for future studies. We start by generating the GW signal in the ideal  case where there are no gaps. This will henceforth be referred to as the `optimal' situation. The gaps are modelled as time dependent step functions in the time domain, which are then used as a window function. Although windowing is done in the time domain, we do not simulate the time domain GW signals directly. This is because it is a time consuming process and, as we shall see in the coming sections, we will be performing a huge number of these calculations for our analysis. Instead, we generate signals in frequency domain, which is faster, and perform an \textit{Inverse Fast Fourier Transform} to get the time domain signal. This signal is then filtered with the window function and transformed back to frequency domain for further analysis. Generation of the optimal signal in the frequency domain is performed using the tools provided in the \texttt{LDC Code} \footnote{https://gitlab.in2p3.fr/stas/MLDC}.

The full distribution of gaps is defined as a collection of many individual gaps. The window function, $w(t)$, for each individual gap can be treated as a time dependent step function which takes the value 0 between $t_{\text{s}}$ and $t_{\text{e}}$, where $t_{\text{s}}$ and $t_{\text{e}}$ are the times at which the gap starts and ends respectively, and is 1 elsewhere. In addition, we also smooth out the window on either ends of the gap with a simple function that allows $w(t)$ to change gradually from 1 to 0 and then back to 1. We do this to avoid any discontinuous changes in the value of $w(t)$, as those lead to spectral leakage when performing Fourier transforms. It does, unfortunately, also lead to greater loss of data. 

The window function of a gap where a smooth transition takes place over a period $t_{\text{tr}}$ can, therefore, be written as %
\begin{eqnarray}
    w(t) =\begin{cases}
    \frac{1}{2}\left( 1 + \cos\Big[\pi(\frac{t - t_{\text{s}} - t_{\text{tr}}}{t_{\text{tr}}})\Big] \right) \nonumber & \text{for\,} t_{\text{s}} - t_{\text{tr}} < t < t_{\text{s}} \nonumber \\
    0 & \text{for\,} t_{\text{s}}< t < t_{\text{e}}\nonumber\\
    \frac{1}{2}\left( 1 + \cos\Big[\pi(\frac{t - t_{\text{e}} - t_{\text{tr}}}{t_{\text{tr}}})\Big]\right) & \text{for\,} t_{\text{e}}< t < t_{\text{e}} + t_{\text{tr}}\nonumber\\
    1 & \text{otherwise}
    \end{cases}
\end{eqnarray}
Many such individual gaps are combined to get a distribution of gaps spanning the mission duration. To simulate the scheduled gaps in data, we form a window function such that there is one such gap of 3.5 hours (7 hours) every week (2 weeks).

To simulate unscheduled gaps, we combine 3 day long gaps by choosing the intervals between the gaps such that the 75\% duty cycle requirement (Sec.~\ref{types_of_gaps}) is complied with. To accomplish this, the interval values are drawn from an exponential distribution with scale parameter ensuring that 25\% of the observation period (when averaged over different realizations of the gap distribution) is taken up by gaps.  Therefore, the interval $\Delta T$ between two consecutive gaps has a probability density function given by
\begin{equation}
    \frac{\mbox{d} p}{\mbox{d} \Delta T} = \lambda e^{-\lambda \Delta T}\,, \qquad \mathrm{where\,\,} \lambda=\frac{1}{9} \mbox{days}
\end{equation}
The intervals can be as small as a few hours and as large as a few weeks. This poses a problem when we try to ensure smooth transitions at the boundaries of the gaps. The amount of spectral leakage in the frequency domain signal depends on the degree of smoothing, which is represented by the time taken to transition (transition time) at the ends of the gap. Large transition times lead to less leakage, but also leads to higher losses of useful data. Empirically, we found that for 3 day gaps, a transition time of 2 days on either side of the gap is a good compromise. The interval between two gaps is, however, not always longer than two days and applying smooth transitions becomes impossible in such cases. To overcome this, we simply combine these type of consecutive gaps into one large gap ($2\times 3 \text{ days} + \Delta T$). An example of a window formed in this way is shown in Fig.~\ref{fig:smooth_windows}. Some of the gaps are clearly longer than others, meaning that these are formed by combining many individual gaps that were separated by intervals of less than what was required for smoothing. It is worth emphasizing that this process conservatively reduces the effective duty cycle and increases the loss in the SNR of the GW signal. Instead of losing 25\% of the data, we lose $\sim$35\%, and the effective duty cycle with this technique is $\sim$ 65\%. As such, this technique of gap smoothing must be applied with caution when working with real data as some signals that were not lost due to instrumental malfunction may be lost while windowing the data.

In Fig.~\ref{fig:smooth_windows}, we show the scheduled gaps distribution and one of the realizations of unscheduled gaps. The smooth transitions at the ends of the gaps are also shown in the right panels. We apply these window functions, $w(t)$, to the signal to simulate the effect of gaps. If $h(t)$ is the optimal signal, then the signal with gaps will be
\begin{equation}\label{eq:window_data}
    h_{\text{w}}(t) = w(t)h(t)
\end{equation}
This time domain signal is then transformed to frequency domain with a simple \textit{Fast Fourier transform}, which is then used for further analysis.

\section{LISA response function and noise model}\label{reponse_and_noise}

In this study, the waveforms are modelled using the non-precessing PhenomD models \cite{2016PhRvD..93d4006H, 2016PhRvD..93d4007K}, while restricting ourselves to the dominant $(2,\pm2)$ modes. It should be noted that the inclusion of higher harmonics may be important for parameter estimation of certain signals (see Sec.~\ref{bayesian_PE}). In general, higher harmonics lead to better estimation of source parameters~\cite{ArunIyer2007,Trias:2007fp,Porter:2008kn,McWilliams:2009bg,2020arXiv200300357M}. The calculated waveforms are treated to the `\textit{full}' LISA response function. A detailed discussion on the construction of this function is beyond the scope of this paper. We will, however, review the basics in brief in this section.

The single-arm interferometry measurements ($y_{slr}$) account for the shift in the frequency of the laser across the link $l$ between the LISA spacecraft $s$ and $r$ due to a passing  GW. We exploit the mode symmetry (for nonprecessing systems) between $h_{22}$ and $h_{2,-2}$ to write the signal in terms of $h_{22}$ only. $y_{slr}$ can then be written as
\begin{equation}
    y_{slr} = \mathcal{T}_{slr}^{22}h_{22}
\end{equation}
where $h_{22} = A_{22}(f)e^{-i\Psi_{22}(f)}$ and $\mathcal{T}_{slr}$ is the single-link transfer function (see \cite{2014LRR....17....2B} for details on the decomposition of GW waveforms in spin-weighted spherical harmonics). A naive calculation of the LISA response function involves costly computations as the spacecraft is not stationary during the lifetime of a signal. Baker and Marsat \cite{2018arXiv180610734M} proposed a novel method for fast calculation of the response function. Following their prescription, the transfer function can be written as
\begin{subequations}\label{eq:transfer}
\begin{equation}\label{eq:transfer1}
    \mathcal{T}_{slr}^{22}(f) = \mathcal{G}_{slr}^{22}(f, t_{f}^{22})
\end{equation}
\begin{eqnarray}\label{eq:transfer_main}
    \mathcal{G}_{slr}^{22}(f, t_{f}^{22}) =&& \frac{i\pi fL}{2}\text{sinc}\left[\pi fL(1 - k\cdot n_l)\right]\nonumber \\
    && \times\text{exp}[i\pi f(L+k\cdot(p_r+p_s))]n_l\cdot P_{22}\cdot n_l\nonumber\\
\end{eqnarray}
\begin{equation}\label{eq:tf_relation}
    t_{f}^{22} = -\frac{1}{2\pi}\frac{d\Psi_{22}}{df}
\end{equation}
\end{subequations}
where $k$ is the GW propagation vector, $L$ = 2.5 Gm is the length of each arm of the LISA constellation, $n_l(t)$ is the link unit vector from spacecraft $s$ to $r$, $p_s(t)$ and $p_r(t)$ are respectively the positions of the spacecraft $s$ and $r$ in the Solar System Barycentre (SSB) frame and $P_{22}$ is the polarization tensor as defined in \cite{2020arXiv200300357M}. The time dependence is implicit in (\ref{eq:transfer}). The position of the spacecraft can also be written as $p_s(t) = p_0(t) + p_s^L(t)$, where $p_0(t)$ is the position of the center of the LISA constellation and $p_s^L(t)$ is the position of the spacecraft $s$ measured in a coordinate system whose origin lies at the center of the LISA constellation. Eq.~\eqref{eq:transfer_main} can then be rearranged for a phase term $\text{exp}[i\pi fk\cdot p_0(t)]$ to appear. This is the Doppler modulation on the phase of the GW signal. $n_l\cdot P_{22} \cdot n_l$ is the projection of the GW on the axes of the interferometer and is, therefore, a manifestation of the antenna pattern function. Both the Doppler modulation and the antenna pattern functions are dependent on time and the sky position of the source. This has interesting consequences for parameter estimation, which are examined in later sections.

Besides, the single link observables are dominated by laser noise. This can be mitigated with Time Delay Interferometery (TDI) \cite{PhysRevD.62.042002, Armstrong_1999, PhysRevD.65.102002, 2005LRR.....8....4T}. In this approach, we calculate  a new set of observables, called TDI A, E and T, which are time delayed linear combinations of $y_{slr}$. Assuming that the noise in the detector arms is uncorrelated, these channels are independent. All further analysis is performed in terms of these TDI observables. %\at{say that $A$, $E$ and $T$ are time delayed liner combinations of the $y_{slr}$.}

\begin{figure}
    \centering
    \includegraphics[width=0.45\textwidth]{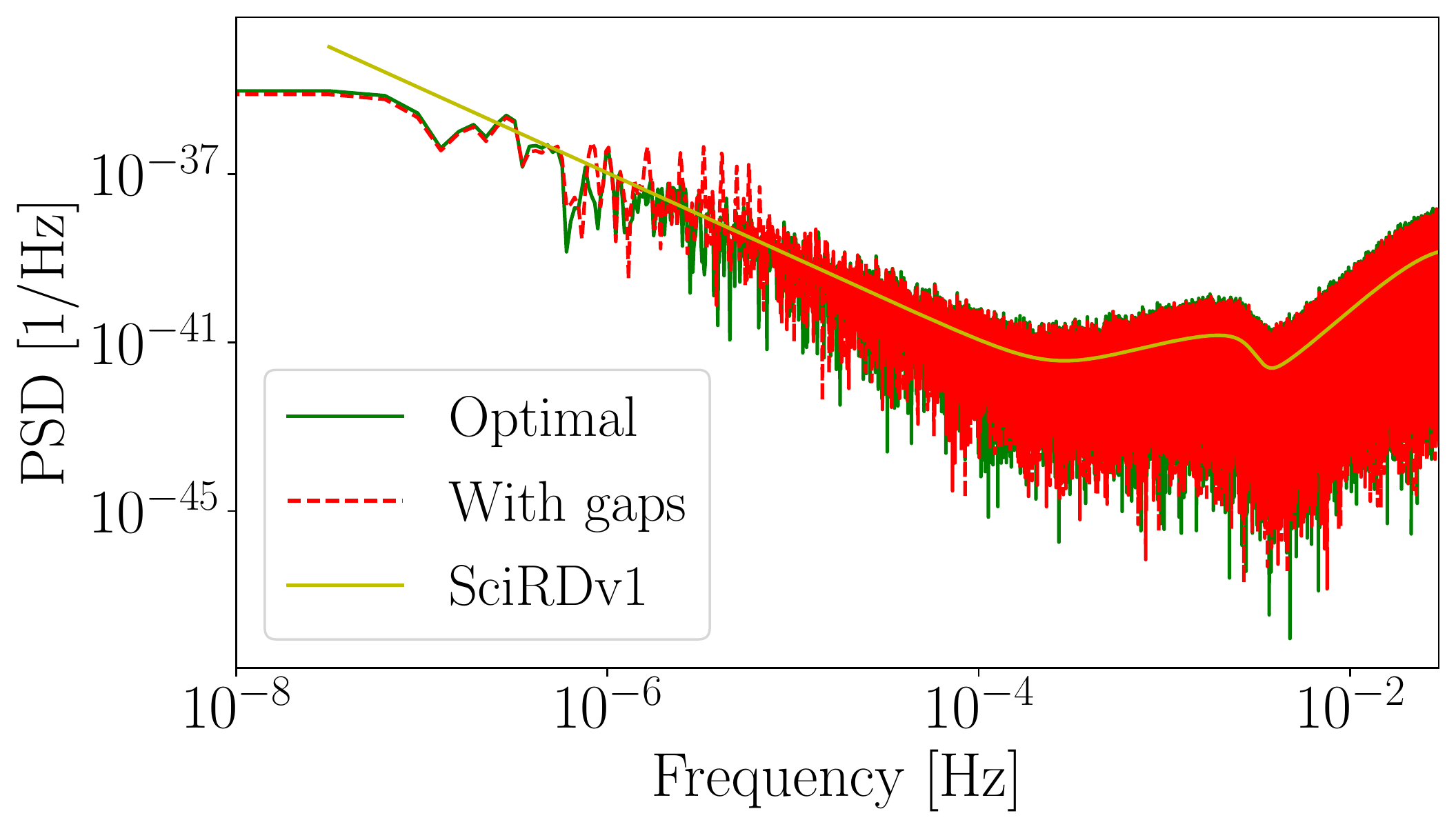}
    \caption{Comparison of the PSD of the noise without gaps and with unscheduled gaps. Presence of scheduled gaps in the data also leads to a similar picture.}
    \label{fig:noise_model}
\end{figure}

\begin{figure*}
\subfloat[Cumulative SNR for the high mass source \label{fig:cumu_snr_heavy}]{%
  \centering
  \includegraphics[width=0.45\textwidth]{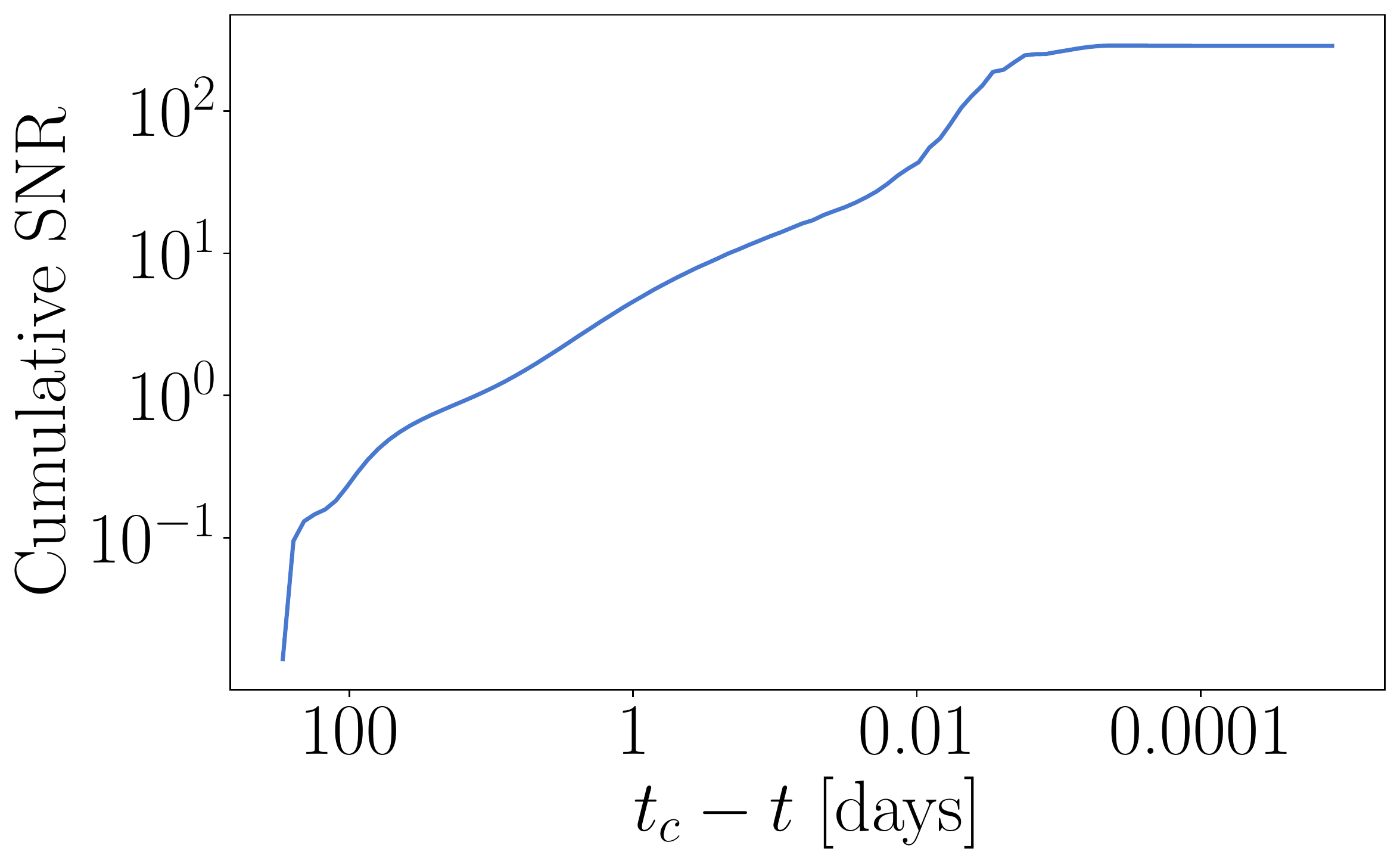}%
}\hfill
\subfloat[Cumulative SNR for the low mass source \label{fig:cumu_snr_low}]{%
  \centering
  \includegraphics[width=0.45\textwidth]{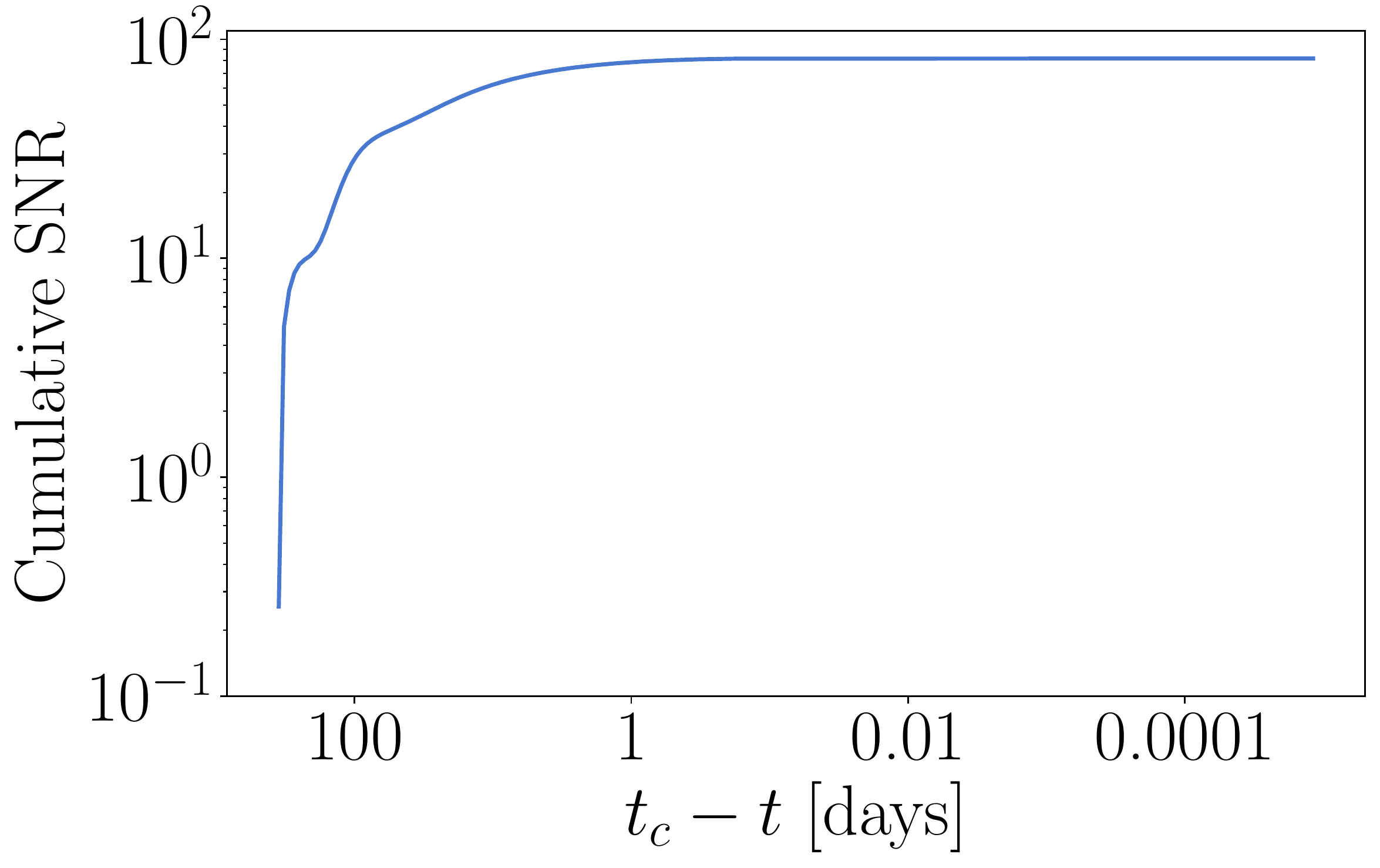}%
}\hfill
\subfloat[SNR loss for a high mass source \label{fig:snr_loss_heavy}]{%
  \centering
  \includegraphics[width=0.45\textwidth]{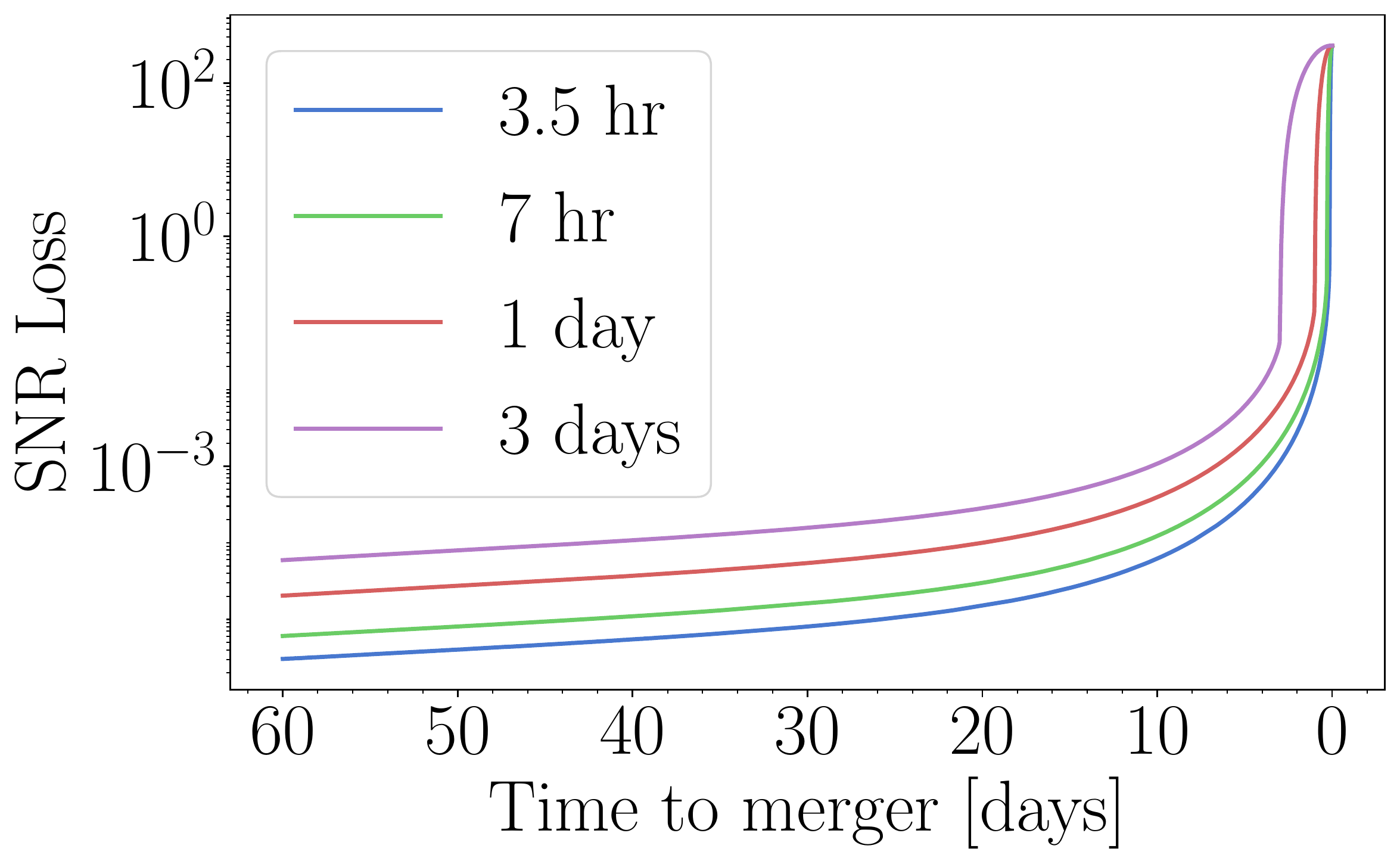}%
}\hfill
\subfloat[SNR loss for a low mass source \label{fig:snr_loss_light}]{%
  \centering
  \includegraphics[width=0.45\textwidth]{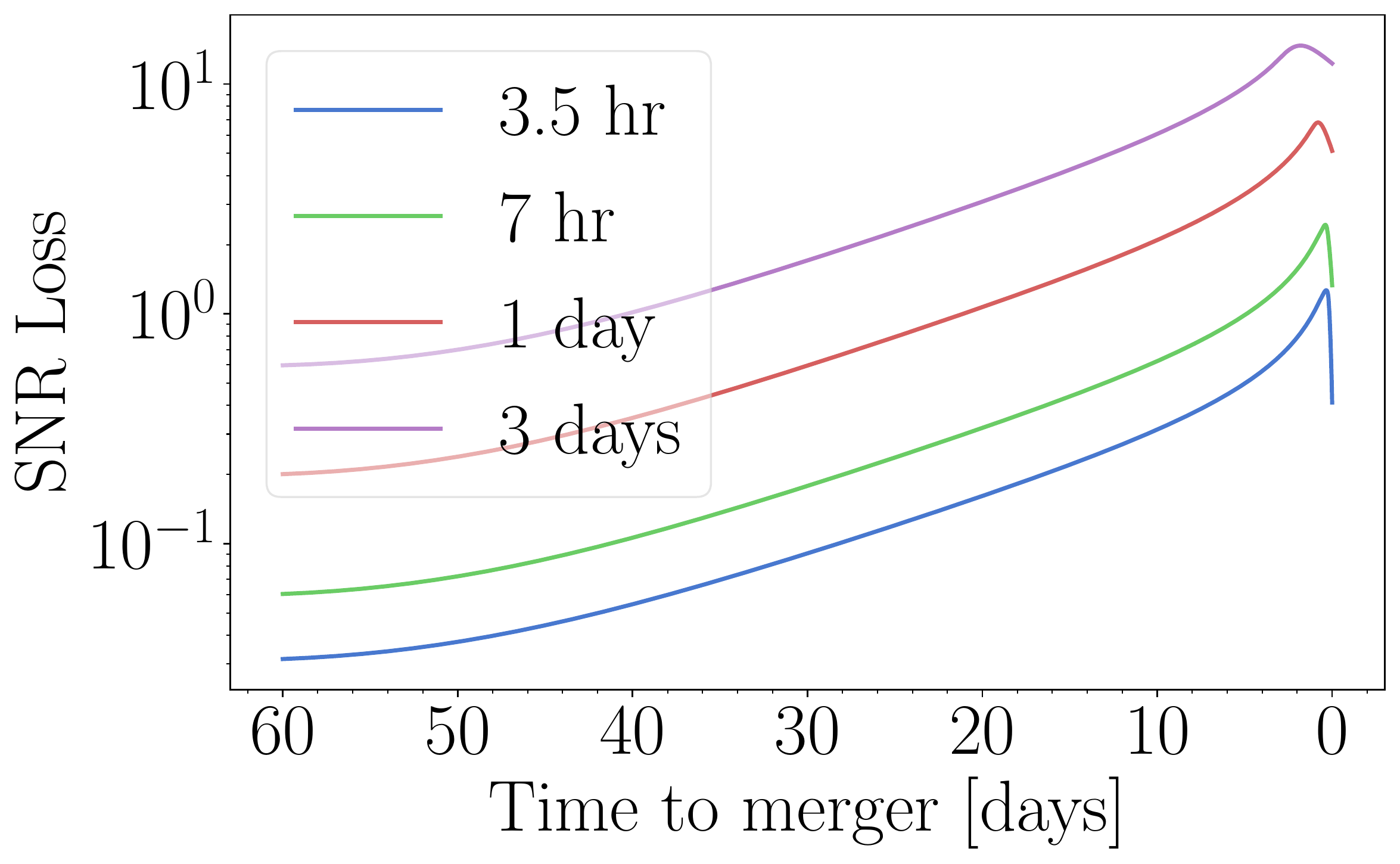}%
}\hfill
\caption{Cumulative SNR of the optimal signal as a function of time from merger (whose time 
is denoted by $t_c$), together with the loss in the SNR of the signal as a function of the distance of the gap from the merger. We consider a heavy mass system (optimal SNR = 308) in (a), (c) and a low mass system (optimal SNR = 81) in (b),(d). While calculating the loss in SNR, we consider individual gaps of different lengths, each shown in different colour. We will use these sources again in Sec.~\ref{bayesian_PE} to study the effect of gaps on parameter estimation.}
\label{fig:snr_loss}
\end{figure*}

The application of smooth transitions at the ends of each individual gap allows us to use the same noise model for both optimal and gapped cases. This can be seen in Fig.~\ref{fig:noise_model}, where we compare the power spectral density (PSD) of the instrumental noise for the two cases. Appreciable differences are seen only at low frequencies ($\sim 10^{-6}$ Hz). Without smoothing, the effect of spectral leakage would force us to use a different noise model. Also seen in Fig.~\ref{fig:noise_model} is the analytical noise model (\texttt{SciRDv1}) \cite{LISA_SciRDv1} in the presence of galactic binary (GB) confusion noise. The LISA signal is expected to contain this GB confusion, which slightly reduces the sensitivity. Further details regarding the noise model are included in Appendix \ref{app:noise}.

\section{Methodology}\label{methodology}

Any kind of gap in the data stream will reduce the SNR of the signal. If $\rho_{\text{opt}}$ and $\rho_{\text{gap}}$ are the SNRs of the signal in the absence and the presence of gaps respectively, then the SNR loss due to the inclusion of gaps is $\rho_{\text{opt}} - \rho_{\text{gap}}$. See section \ref{SNR_des} for further details on the SNR calculation. The magnitude of this loss depends on the placement and type of the GW source being considered. In this paper, we concentrate on MBHBs. Depending on the binary mass, the signal can be transient or long lasting. Highly massive systems ($\gtrsim10^5 M_\odot$) are generally transient, with most of the signal accumulating near the merger, in the highly sensitive part of the LISA band. The SNR of the low mass systems ($\sim 10^3 - 10^5 M_\odot$) is more evenly distributed over the duration of the signal, with mergers taking place at relatively high frequencies, in a part of the LISA band where the sensitivity is low. In fact, we may only observe the inspiral phase for some of these systems.  

With gaps in MBHB signals, the SNR loss increases as the gap falls closer to the merger, and is maximum when the merger happens within a gap. This is illustrated in Fig.~\ref{fig:snr_loss}, where we plot the loss in the SNR of a transient (${\rm SNR}_{\rm optimal}$ = 308, Fig.~\ref{fig:snr_loss_heavy}) and long lasting (${\rm SNR}_{\rm optimal}$ =81, Fig.~\ref{fig:snr_loss_light}) MBHB signal for different positions (and durations) of the gap. As the gap gets close to the merger, the loss increases rapidly for the transient signal and is negligible for gaps located far from coalescence. For a long lasting signal, however, the loss in SNR is much more gradual, as the SNR is more evenly distributed. As the merger takes place at  higher frequencies, where the sensitivity of LISA is low, the contribution to the cumulative SNR from the merger-ringdown phase is lower than for a transient signal. This results in a decrease in loss of SNR very close to the merger (observed in Fig.~\ref{fig:snr_loss_light}).  

Loss of SNR  is expected to hamper parameter estimation and clearly depends on the position of the gaps relative to the merger. Because of this, we decided that instead of focusing only on parameter estimation with individual sources, we would look at synthetic catalogs of MBHB sources, calculated with different astrophysical models. This would provide us with a more comprehensive picture of the impact of gaps, as these catalogs represent expected astrophysical populations that LISA may observe. 

% more about the catalogs here.

We consider the GW signal due to each source in the catalogs and simulate gaps by following the process outlined in Sec.~\ref{data_prepare}. For the scheduled gap distribution, we consider two representative scenarios -- 3.5 hour gaps every week and 7 hour gaps every two weeks during the  observation period. As mentioned before, the placement of these gaps can be decided beforehand. Both the scenarios being viable alternatives for this purpose, we want to check if there is any evidence of one of these being better than the other from the perspective of data analysis. For the unscheduled gaps, we build the window function by combining individual gaps, each of which is 3 days long, while ensuring smooth transitions (see Sec.~\ref{data_prepare}), and apply it to the data using Eq.~\eqref{eq:window_data}. We consider the scheduled and unscheduled gaps separately to facilitate easy comparison between the two. 

With both the \textit{optimal} and \textit{gapped} data, we compute the SNRs, followed by the Fisher Information Matrices (FIMs) for each source. Comparing the SNRs tells us how much of the signal is lost due to the gaps. We then attempt to estimate the parameters for some representative sources.

\subsection{Signal to Noise Ratio}\label{SNR_des}

The SNR is a measure of the strength of the GW signal in the data. It is calculated as
\begin{equation}\label{eq:snr}
    \rho^2 = \left(h|h\right)
\end{equation}
where $\rho$ is the SNR of the signal and $h$ denotes the GW strain. The inner product is defined as
\begin{equation}\label{eq:inner_product}
    \left(a|b\right) = 4\text{Re}\int_{0}^{\infty}\frac{\Tilde{a}^*(f)\Tilde{b}(f)}{S_n(f)}\, \mathrm{d}f
\end{equation}
where $S_n(f)$ is the PSD of the LISA noise. While calculating the SNR in the presence of gaps, $h$ is replaced by $h_\text{w}$ (see Eq.~\eqref{eq:window_data}). We work with the TDI observables A and E, which form two independent data streams (note that T has very low signal content). The total SNR of the data is thus the square root of the sum of the SNRs of these data streams, i.e.
\begin{equation}\label{eq:SNR_eq}
    \rho_{\mathrm{total}}^2 = \rho_{A}^2 + \rho_{E}^2
\end{equation}
where $\rho_{\mathrm{total}}$ is the total SNR, while $\rho_A$ and $\rho_E$ are the SNRs of the A and E streams respectively. The LISA sensitivity is described in Sec.~\ref{reponse_and_noise} and Appendix \ref{app:noise}. The SNR will decrease with the introduction of gaps. The magnitude of this loss is an indicator of how much the parameter estimation will be impacted.

\subsection{Fisher Information Matrix}\label{FIM_des}

Obviously, if any data is lost, the errors in the recovered parameters will be larger than those computed for the optimal case. The difference in the errors highlights the effect of gaps. Large differences mean that data analysis is adversely affected. The ideal way to calculate these errors would be to estimate the posterior distributions with Bayesian techniques, for each GW source. Bayesian parameter estimation, however, is a computationally expensive process and it is not feasible to run any of these techniques on all the sources in our simulated catalogs. An alternative is to calculate the FIM. The inverse of the FIM gives the Cramer-Rao lower bound, which is a limiting value of the covariance of the source parameters in the high SNR limit \cite{2008PhRvD..77d2001V}. It should be noted that the Cramer-Rao  bound is only
approached at high SNR. For low SNR signals originating from low mass systems, the error estimates from the FIM are, therefore, optimistic \cite{2008PhRvD..77d2001V, PhysRevD.57.4588}. This does, however, provide a good baseline for further studies, and allows us to keep the computational cost reasonable, especially since we will be considering a large number of sources (c.f. Sec. \ref{astro_catalogs}). This allows us to set bounds on the best achievable accuracy of parameter estimation techniques. 

The FIM elements are calculated as 
\begin{equation}\label{eq:fisher_definition}
    F_{ij}(\theta) = (\partial_ih|\partial_jh)\Big|_{\theta}
\end{equation}
where $\partial_i$ denotes the derivative with respect to parameter $\theta_i$ and $h$ is the GW signal in the frequency domain. In the presence of gaps, $h$ is obtained from the Fourier transform of $h_\text{w}(t)$ in (\ref{eq:window_data}).

Implementing the scheme defined by (\ref{eq:fisher_definition}) with numerical differentiation is non-trivial. The process of numerical differentiation, and consequently the FIM elements, are extremely sensitive to step sizes used for the different parameters ($\theta_i$). Ideally, one would vary the step sizes gradually and compare the corresponding eigenvalues of the FIM. For the appropriate value of step size, the eigenvalues would vary negligibly. This method is time consuming as one has to repeat this for each source. Alternatively, one can choose a differentiation scheme with a high accuracy. It is worth noting that this increases the time it takes to calculate FIM for each source. The choice of the order of accuracy must strike a balance between efficiency and precision. 
In our analysis, we use a fourth order central finite difference scheme to calculate the derivatives. We further reinforce our calculations of the derivatives with  Richardson extrapolation. 

We use this framework to calculate the Cramer-Rao lower bounds on the errors for the source parameters of the GW signals corresponding to the various catalogs. We then choose a few representative cases for parameter estimation within a Bayesian framework. 

\subsection{Parameter estimation}\label{PE_des}

Realistically, the data recorded by LISA, $\textbf{d}$, will be in the form of a GW signal (\textbf{h}) superposed with noise (\textbf{n}), $\textbf{d} = \textbf{h} + \textbf{n}$. For estimation of source parameters, let us work in a Bayesian framework. Bayes theorem states that
\begin{equation}\label{eq:bayes_thm}
    p\left(\theta|d \right) = \frac{p\left(d | \theta \right)p\left(\theta \right)}{p\left( d\right)}
\end{equation}
where $ p\left(\theta|d \right)$ is the posterior probability distribution, $p\left(d | \theta \right)$ is the likelihood, $p\left(\theta \right)$ is the prior probability distribution and $p\left( d\right)$ is the evidence. In the subsequent sections, the likelihood function will be referred to as $\mathcal{L}(\theta)$. Evidence is used to compare the viability of different signal and noise models. The evidence can be considered to be a normalization constant as the noise and GW signal models will remain the same throughout the study. 

Assuming the noise to be stationary and Gaussian, the likelihood is
\begin{equation}\label{eq:logl}
    \mathcal{L}(\theta) = \mathcal{A}\cdot\text{exp}\big[-\frac{1}{2}\left( \textbf{d} - \textbf{h}(\theta)|\textbf{d} - \textbf{h}(\theta)\right)\big]\\
\end{equation}
where $\mathcal{A}$ is the normalization constant and the inner product is defined in Eq.~\eqref{eq:inner_product}. In practice, we will actually calculate the logarithm of $\mathcal{L}(\theta)$. This quantity will be referred to as loglikelihood in the rest of the paper. Furthermore, we fix the noise at zero $(n=0)$ when simulating the data. Under the assumption of stationary and Gaussian noise, this corresponds to an average over all possible noise realizations~\cite{Rodriguez_2014, Sampson_2013PRD, Nissanke_2010}. The effect of a noise realisation to parameter estimation, is an overall shift of the posterior distribution, while the posterior widths (or structure in general) remains the same. In this work, we choose to adopt the noiseless case in favour of faster computations and in order to simplify the process. Otherwise, a Monte Carlo over noise realisations would be necessary. Moreover, the systems will be analysed individually, i.e. we assume that the data contains GW signals from only one source at a time. As stated before, the TDI A, E and T are noise-independent channels and the total loglikelihood will be the sum of loglikelihoods for these three channels. The T channel contains negligible information about the signal for noiseless data, and can be ignored in order to speed up calculations. 

Calculation of the loglikelihood at a point in the parameter space is a computationally expensive process when one includes the gaps. The gaps are applied in the time domain (see Sec.~\ref{data_prepare}). The signal in the A and E channels are transformed to the time domain, where the gaps are applied and are then transformed back to the frequency domain, for calculation of the loglikelihood using Eq. (\ref{eq:logl}). Robust sampling techniques are  needed to achieve convergence of the posterior samples. 

The parameters can be broadly classified into two categories - intrinsic and extrinsic. The intrinsic parameters include the chirp mass, $M_c = (m_1m_2)^{3/5}/(m_1+m_2)^{1/5}$, the mass ratio, $q = \frac{m_1}{m_2}$ (where $m_1$ and $m_2$ are the masses of the primary and secondary black hole respectively), spins, $\chi_1$ and $\chi_2$, and time to coalescence, $t_c$, of the system. These five parameters describe the inherent properties of the source, irrespective of the relative orientation of the observer. The extrinsic parameters include the position in the sky as seen in the SSB frame ($\lambda$ and $\beta$), the luminosity distance to the source, $D_L$, the inclination of the orbital angular momentum of the source with the line of sight, $\iota$, and the polarization angle, $\psi$. In addition, the phase at coalescence also acts as a parameter. 

Previous studies on the parameter estimation of MBHB signals have mostly relied on using simplified signal models, with an approximate version of the instrument response. Historically, the Fisher matrix approximation was preferred over full Bayesian techniques owing to the computationally intensive nature of the latter \cite{2004PhRvD..70d2001V, 2006PhRvD..74b4025A, 2005PhRvD..71h4025B, 2006PhRvD..74l2001L}. The main limitations of the  Fisher matrix approximation  is that it does not resolve secondary maxima in the posterior distributions, and it allows for computing the errors in the parameter values without estimating their true values. Moreover, the inverse of the Fisher matrix gives a tight lower bound on the errors only in the high-SNR limit \cite{PhysRevD.57.4588}. At very low SNRs, the error estimates obtained from the Fisher matrix approach are usually optimistic. In these cases, a full Bayesian analysis is the only option. The Mock LISA Data Challenges
and the ongoing
LISA Data Challenges provide simulated LISA data, 
 with the goal of
 generating interest in the various aspects of LISA data analysis. These challenges utilize simple models for the waveforms and response, but also provide Markov Chain Monte Carlo (MCMC) tools for the analysis \cite{2010CQGra..27h4009B}. With this easy availability of computational facilities and development of robust sampling techniques, there have indeed been some studies in the last few years that explore the full parameter space of MBHB signals \cite{2020arXiv200300357M, 2020PhRvD.102b3033K, PhysRevD.101.124008}.

In order to navigate the parameter space and draw posterior samples, we choose the publicly available sampler \texttt{PTMCMCSampler} \cite{justin_ellis_2017_1037579}, which implements a Parallel Tempering MCMC (PTMCMC) \cite{PhysRevLett.57.2607, B509983H} scheme. PTMCMC has been observed to be quite efficient at sampling positions in the parameter space that are difficult to access with traditional Metropolis-Hastings method \cite{10.2307/2280232, doi:10.1063/1.1699114, 10.2307/2334940}. It has been successfully applied to GW signal analysis \cite{2009PhRvD..80f3007L, 2020arXiv200300357M, 2020PhRvD.102b3033K}. The sampler itself has been previously tested in the LISA Data Challenges, where it performed well. Although a full discussion on PTMCMC is beyond the scope of this paper, we present a small review of the technique.

PTMCMC is an extension of the Metropolis Hastings technique that increases the efficiency of the search for  global maxima. It does so by means of multiple chains that explore the parameter space in detail, each sampling at a different ``temperature''. Essentially, the likelihood functions becomes $\mathcal{L}(\theta)^{1/T_i}$, where $T_i$ is the temperature of the chain. The chains at higher temperatures sample portions of parameter space that are effectively out of reach for Metropolis-Hastings. The information gathered by the chains at different temperatures are exchanged by means of randomized swaps of parameter values. This exchange is quantified by a modification of the Metropolis Hastings ratio 
\begin{equation}
    H_{i \leftrightarrow j} = \text{min}\left(1, \left[\frac{\mathcal{L}(\theta_i)}{\mathcal{L}(\theta_j)}\right]^{\beta_j - \beta_i}\right)
\end{equation}
where $\beta_i = 1/T_i$. These exchanges usually take place between nearby temperatures. The $\beta = 1$ chain corresponds to the true likelihood function and samples the true posterior distribution. The exchanges that it goes through with the ``hotter'' chains allow it to rapidly  explore the parameter space without getting stuck in any primary or secondary maxima.

Even with the robustness of PTMCMC, the long computation times involved in loglikelihood calculations are a cause of concern. One of the ways to reduce this is to consider an observation duration ($T_{obs}$) of 1 year, instead of the actual LISA mission duration of 4 years. This can be done for MBHB signals, especially the high mass systems, as most of the signal is accumulated near the merger and they spend less than one year in band. The application of gaps is performed in the time domain, as stated above. Fast Fourier transforms have cost of order \textit{O}(\textit{N}log\textit{N}), where $N$ is the size of the data array. The cadence of the LISA signal is set to 10 seconds. With this cadence fixed, increasing $T_{obs}$ increases the time needed for each likelihood calculations, as the number of data points involved increases. We avoid this by choosing a duration of 1 year. Even with this choice of observation duration, the time needed for the loglikelihood computation of the gapped cases exceeds its optimal counterpart by a factor of 5. It would, however, be better to use the actual mission duration if the computational bottlenecks can be dealt with. For low mass systems with longer inspirals, having a longer observation duration will improve the recovered posteriors. We are, however, interested in comparing the posterior distributions in the presence and absence of gaps and $T_{obs} = $ 1 year is sufficient for this purpose, as we analyse both the optimal and gapped signals for the same duration.

We estimate the parameters of two different systems using \texttt{PTMCMCSampler} for both optimal and gapped cases. These systems and their posterior distributions are described in Sec.~\ref{bayesian_PE}.

\section{Astrophysical Catalogs}\label{astro_catalogs}

To describe the expected population of MBHBs detectable by LISA, we use the 
semi-analytic galaxy formation model of~\cite{2020ApJ...904...16B} (based on \cite{Barausse:2012fy},
with successive developments described in \cite{Sesana:2014bea,Antonini:2015sza,Antonini:2015cqa,PhysRevD.93.024003,Barausse:2017uyr,Bonetti:2018tpf} and in \cite{2020ApJ...904...16B} itself). The model follows the evolution, along dark-matter merger trees, of the galactic baryonic  structures,  including a diffuse chemically primordial intergalactic medium, which accretes onto dark-matter halos either along cold flows (at high redshifts and/or small systems) or by getting shock-heated to the halo virial temperature (in large low-redshift systems)~\cite{Dekel2006,Cattaneo2006,Dekel2009}; 
a cold interstellar medium (with spheroidal and/or disk-like geometry), which
harbors star formation (including supernova explosions, which exert a feedback on star formation and which chemically enrich the interstellar medium);
stellar disks and spheroids; nuclear star clusters; and massive black holes,
which grow by accretion and mergers from high redshift seeds, and which
inject energy into the surrounding gas through jets, disk winds and radiation -- a process typically referred to as feedback from Active Galactic Nuclei (AGNs).

The properties of the population of MBHBs detectable by LISA depend on the detailed evolution of the dark-matter and baryonic structures, to which they are inextricably linked (through accretion, mergers, feedback, etc). Calibration to a number of observations (see e.g.~\cite{Barausse:2012fy,Sesana:2014bea,Antonini:2015sza,Antonini:2015cqa,Barausse:2017uyr,Guepin:2017abw,2020ApJ...904...16B}) allows  for reducing the uncertainties resulting from this connection. There are however aspects of the model for which observational constraints are looser, and which have a significant impact on the MBHB population, for instance: \textit{(i)} the mass function of the high-redshift ``seeds'' from which massive black holes grow (see e.g.~\cite{Latif:2016qau} for a review); 
\textit{(ii)} the effect of supernova feedback on the growth of  massive black holes via accretion (which may be curtailed by supernova winds in systems with shallow potential wells~\cite{Habouzit2017}); and \textit{(iii)} the distribution of the time ``delays'' between the merger of two dark-matter halos and the merger of the corresponding MBHB (see e.g.~\cite{Colpi2014} for a review).

To bracket the uncertainties due to these physical effects, we will consider here a suite of eight models among those described in~\cite{2020ApJ...904...16B}, and namely those singled out in~\cite{Barausse:2020gbp}. In more detail, we consider ``light-seed'' models, in which
massive black holes form from the remnants of population-III stars at $z\gtrsim 15$, with masses of a few hundreds of $M_\odot$~\cite{Madau2001}; and ``heavy-seed'' models, where
massive black holes also form at high redshift  $z\gtrsim 15$, but with much larger masses $\sim 10^4$--$10^5 M_\odot$~\cite{Volonteri2008} (see also \cite{2019RPPh...82a6901M} for a review of heavy-seed formation scenarios). 
Both seed models are combined with accretion prescriptions
neglecting supernova feedback (``noSN'' models)
on the accretion flow (though feedback on star formation is always included); and with prescriptions accounting for 
the possible quenching of accretion due to SN winds in small systems (``SN'' models)~\cite{Habouzit2017}. As for the delays distribution,  we consider ``short-delays'' models,
where the delays at MBHB separations of hundreds of pc  are neglected (as was done in most of the prior literature), as well as more realistic ``delays'' models where those contributions are properly taken into account~\cite{Dosopoulou2017,Tremmel2018}. (Note however that the delays
due to dynamical friction between the halos and the galaxies~\cite{Binney2008,Taffoni2003,Boylan-Kolchin2008}, and due to
stellar hardening~\cite{Quinlan1996,Sesana2015,Vasiliev2015}, gas-driven migration~\cite{MacFadyen2008,Cuadra2009,Lodato2009,Roedig2011,Nixon2011,Duffel2019,Munoz2019} and triple/quadruple massive black hole interactions~\cite{Bonetti:2018tpf,Bonetti2018a} are included in all of our  models, as they are comparatively better understood than the delays at separations of hundreds of pc).

Our suite of eight models is then obtained by combining these three binary options
(light vs heavy seeds; noSN vs SN; short-delays vs delays). The redshift, mass, mass ratio and signal-to-noise ratio distributions of the eight populations differ significantly from one another, as discussed in detail in~\cite{2020ApJ...904...16B,Barausse:2020gbp}. In particular, the light-seed SN models typically predict less massive binaries 
(with the most realistic model, the SN-delays one, featuring a relatively broad mass distribution, but with a peak at total MBHB masses of $\sim 10^3 M_\odot$)
and lower signal-to-noise ratios (typically $\lesssim 100$), which may be challenging for LISA detection and parameter estimation, especially in the presence of gaps.

\section{Impact of gaps}\label{impact_of_gaps}

As a result of gaps in the data, a part of the signal is lost. As a rule of thumb, this can be quantified by the loss in the SNR of the signal. Typically, the greater the difference between the SNRs of the signal for optimal and gapped cases, the greater will be the impact of gaps on subsequent parameter estimation. Studying the SNR is therefore a natural starting point. As can be seen from Fig.~\ref{fig:snr_loss}, the placement of gaps affects the magnitude of data loss. We consider a few catalogs of sources for different astrophysical models (see Sec.~\ref{astro_catalogs}), which will allow us to make statistically relevant predictions. This is followed by the calculation of the FIM, which will give us an idea about the errors incurred due to the different types of gaps for the various astrophysical populations under consideration. We choose the sky position and the orientation of the source randomly from an isotropic distribution. We do not average over the sky for any of the subsequent calculations. We consider both the scheduled and unscheduled gaps scenarios (see Sec.~\ref{types_of_gaps} and \ref{data_prepare}), and apply them to the data separately. Our findings are described in the following sections.

\subsection{Loss in SNR}

The SNR is calculated using Eq.~\eqref{eq:SNR_eq}. We calculate the total SNR of the data in the presence and absence of the different types of gaps, for all the catalogs under consideration. A high SNR usually implies narrower posterior distributions. If the SNR is too low, estimation of source parameters corresponding to the signal may not be possible. We take $\rho_{th} = $ 8 as the SNR detection threshold. It should be noted that we do not focus on the problem of detection in this study. The detection of low SNR signals, particularly like the ones seen for the light seed models, would indeed be difficult. We, however, assume that the data stream has already been determined to contain a GW signal. 

Any signal whose SNR is greater than $\rho_{th}$ is considered as potentially detectable. Each simulated MBHB catalog contains information about mergers that take place over a duration of 300 years, divided into smaller periods of 1 year. The number of potential detections are averaged over a period of 4 years (LISA nominal mission duration). Table \ref{tab:detection_rates} shows the total number of sources and the number of potential detections over a 4 year period for the optimal and the gapped cases.

 \begin{table}[h!]
  \centering
  \caption{Total number of sources and potential detections expected in $4$ yr of observation with LISA for all the models under study.}
  \begin{tabular}{|c|c|c|c|c|}
    \hline 
    
    \multirow{2}{*}{Model}  &  \multicolumn{2}{c|}{\textbf{Light Seed}} & \multicolumn{2}{c|}{\textbf{Heavy Seed}}   \\
    \cline{2-5}
                    & Total & Detected & Total & Detected \\
    \hline \hline
    \multicolumn{2}{|l|}{\emph{\textbf{Optimal}}} \\ \hline
    SN-Delays & 47 & 6 & 24 & 24 \\
    \cline{1-5}
    noSN-Delays & 190 & 101 & 10 & 10  \\
    \cline{1-5}
    SN-shortDelays & 178 & 19 & 1255 & 1255 \\
    \cline{1-5}
    noSN-shortDelays & 1183 & 205 & 1274 & 1274  \\\hline\hline
    
    \multicolumn{2}{|l|}{\emph{\textbf{Scheduled gaps (3.5 h/w)}}} \\ \hline
    SN-delays & 47 & 5 & 24 & 24 \\
    \cline{1-5}
    noSN-delays & 190 & 91 & 10 & 10   \\ 
    \cline{1-5}
    SN-shortDelays & 178 & 13 & 1255 & 1198 \\
    \cline{1-5}
    noSN-shortDelays & 1183 & 172 & 1274 & 1218  \\\hline\hline
    
    \multicolumn{2}{|l|}{\emph{\textbf{Scheduled gaps (7 h/2w)}}} \\ \hline
    SN-delays & 47 & 6 & 24 & 24 \\
    \cline{1-5}
    noSN-delays & 190 & 96 & 10 & 10   \\ 
    \cline{1-5}
    SN-shortDelays & 178 & 17 & 1255 & 1216 \\
    \cline{1-5}
    noSN-shortDelays & 1183 & 188 & 1274 & 1238  \\\hline\hline
      
    \multicolumn{2}{|l|}{\emph{\textbf{Unscheduled gaps}}} \\ \hline
    SN-delays & 47 & 2 & 24 & 19 \\
    \cline{1-5}
    noSN-delays & 190 & 51 & 10 & 8   \\
    \cline{1-5}
    SN-shortDelays & 178 & 5 & 1255 & 697 \\
    \cline{1-5}
    noSN-shortDelays & 1183 & 88 & 1274 & 711  \\\hline\hline

    \hline
  \end{tabular}
  \label{tab:detection_rates}
 \end{table}

As is evident from Table \ref{tab:detection_rates}, the scheduled gaps have negligible impact on the number of sources that can be potentially seen by LISA, whereas the unscheduled gaps lead to much lower detection rates. This is due to the difference between their respective duty cycles. The scheduled gaps result in a duty cycle of $\sim$ 95\% for the mission duration, as opposed to $\sim$ 65\% (effectively) for the unscheduled gaps. For the transient signals (heavy seed catalogs), this means that it is much less likely for the merger to be completely lost due to scheduled data gaps than it is for unscheduled gaps. In fact, as we shall see from the calculation of the FIM, the impact of scheduled gaps is statistically insignificant. We further note that for the long lasting signals (light seed catalogs), only a small fraction of the sources can be detected, even for the optimal case (see \cite{2020ApJ...904...16B,Barausse:2020gbp}). This is because the masses of these sources are typically low ($10^2 - 10^5 M_\odot$), and they merge outside or in the high-frequency end of the LISA  band, which leads to lower SNRs, even in the optimal case. The presence of gaps further decreases the SNR, with unscheduled gaps being especially dangerous.

We further observe from from Table \ref{tab:detection_rates} that the number of potential detections is slightly higher for the scenario with one 7 hour gap every two weeks than for the scenario with a 3.5 hour gap every week. This can be attributed to the windowing process. Even though both cases lead to a total loss of 7 hours worth of data every two weeks, the data loss due to smooth transitions is more for the 3.5 hour case. The number of gaps for the 3.5 hour gap scenario is twice the number of gaps for the 7 hour gap scenario, leading to double the amount of data lost due to transitions for the former compared to the latter. This leads to the observed difference between the number of detectable sources. The relative difference between the two scenarios, however, is small for most of the catalogs that we considered. 

\begin{figure}
    \centering
    \includegraphics[width=0.5\textwidth]{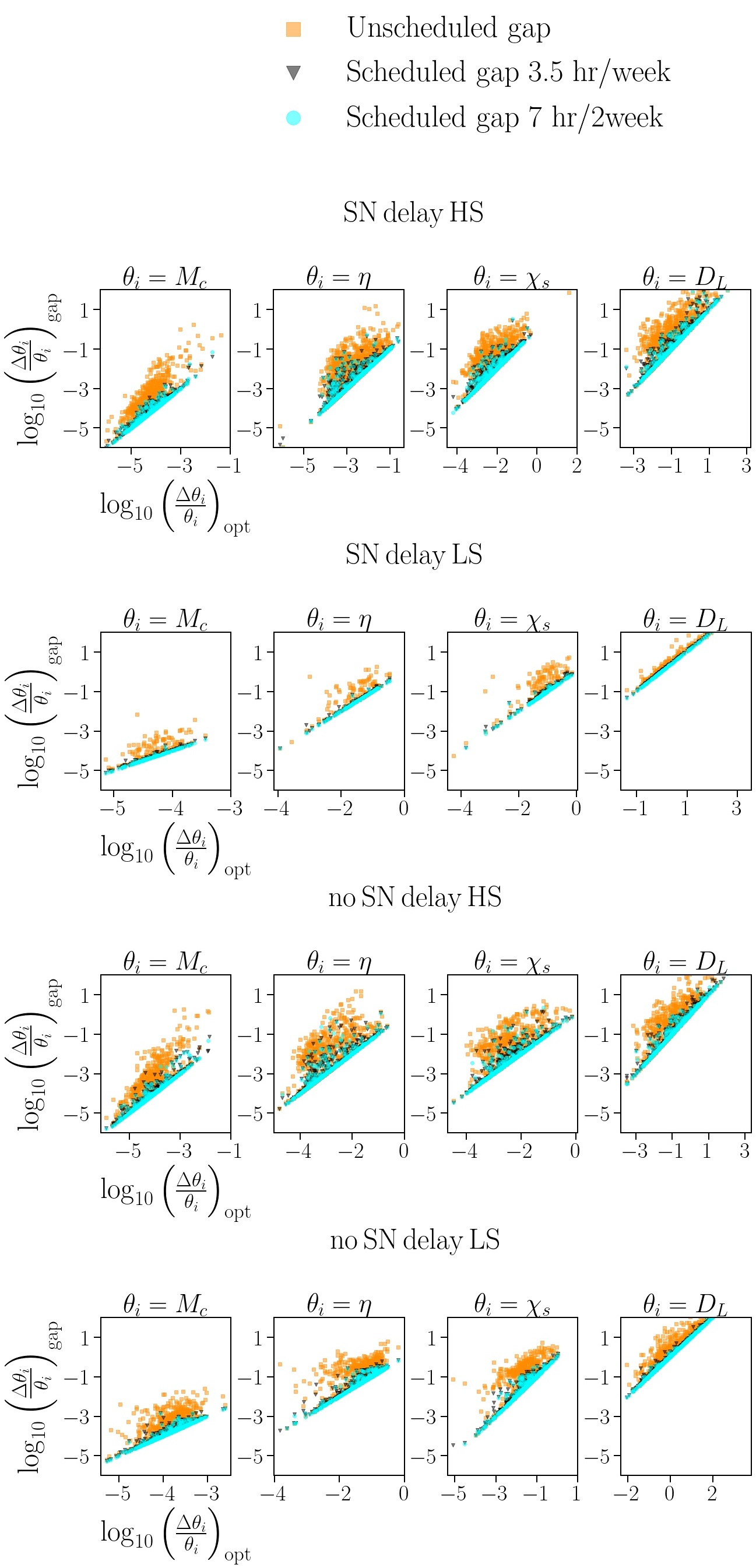}
    \caption{Comparison of relative errors on chirp mass, symmetric mass ratio, effective spin and distance for the optimal and gapped cases, for the sources in the `delays' catalogs from Table \ref{tab:detection_rates}. Along the x-axis, we plot the logarithm of relative errors for the optimal case, while along the y-axis we plot the same for the different gap scenarios. 
    }
    \label{fig:rel_err}
\end{figure}
\begin{figure}[htp]
    \centering
    \includegraphics[width=0.4\textwidth]{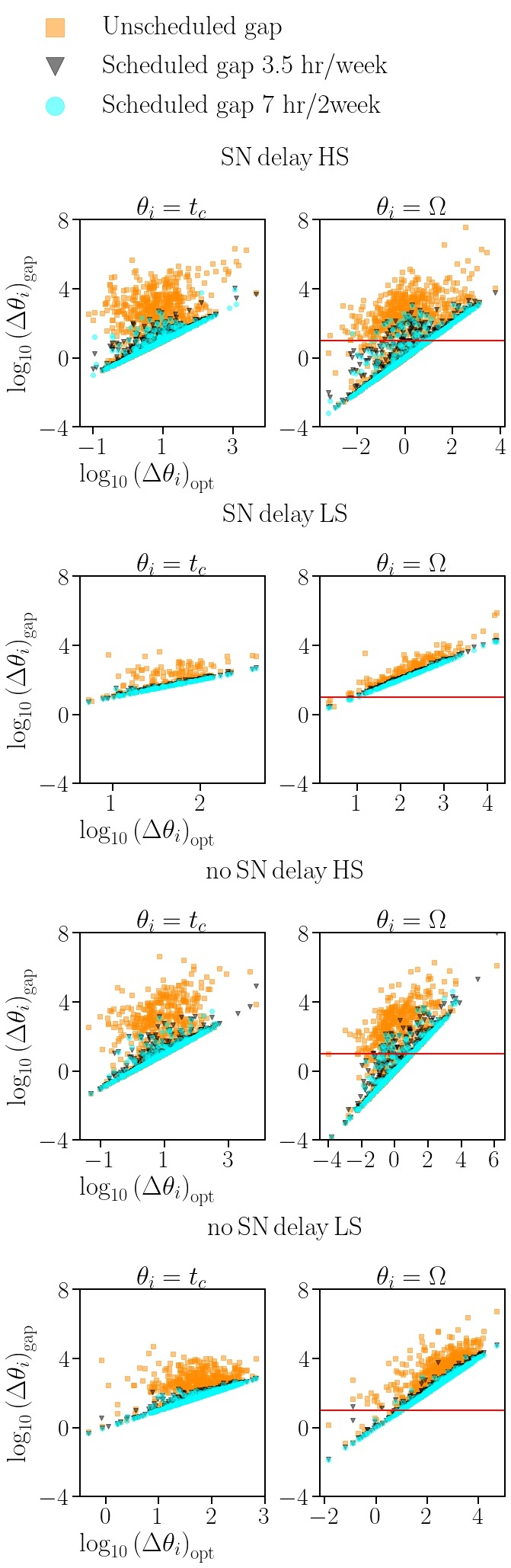}
    \caption{Same as Fig.~\ref{fig:rel_err}, but for the merger time and sky localization.
    }
    \label{fig:rel_err_tc_omega}
\end{figure}

\begin{figure*}[htp]
    \centering
    \includegraphics[width=\textwidth]{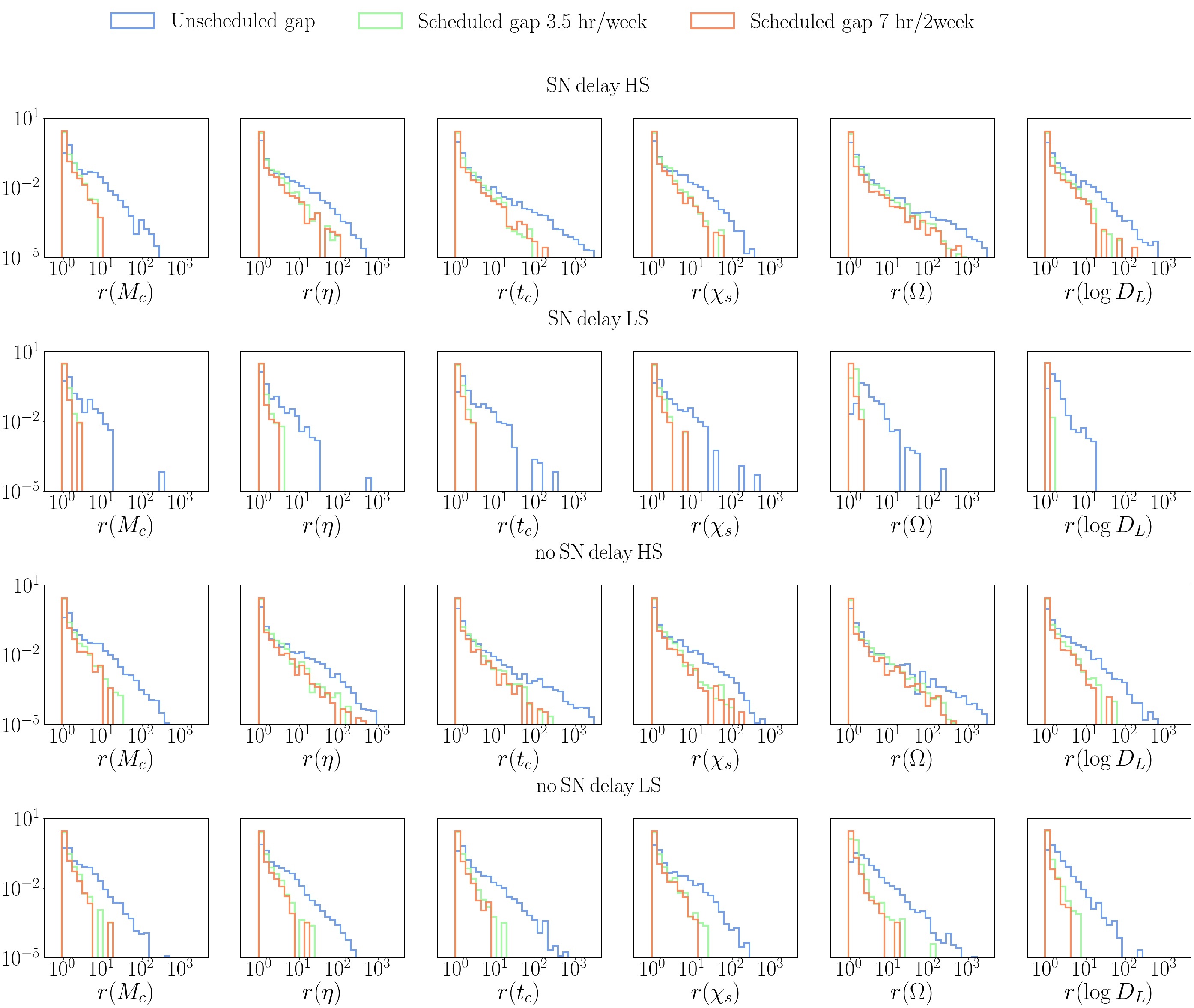}
    \caption{Probability distribution of the ratio of errors $r(\theta)$ between gapped and optimal cases for the sources in `delays' catalogs, from Table \ref{tab:detection_rates}}
    \label{fig:relopt_err}
\end{figure*}

These observations are indeed a cause for concern, especially for the detection of sources from the models that take into account supernova feedback on black hole accretion and realistic delays. This, unfortunately, is an unavoidable consequence of losing a large chunk of the signal due to the gaps. A possible way to mitigate this problem is to use data augmentation techniques. Baghi et al. \cite{PhysRevD.100.022003} have examined one such method. It is worth noting that data augmentation will not make up for the SNR loss in the gaps. It will, however, avoid using smooth windowing, recovering the SNR lost in the transitions. It remains to be seen if these techniques can be optimized for the large gaps considered in this study.

\subsection{Error estimates}\label{errors_fim}

FIMs provide a simple framework to find an estimate of the lower bound of the errors that we expect to find with a full Bayesian analysis. We calculate the FIM from Eq. \eqref{eq:fisher_definition}.
The variance of the unbiased estimator, $\hat{\theta}_i$, of the parameter $\theta_i$ satisfies the Cramer-Rao bound
\begin{equation}
    \text{var}(\hat{\theta}_i) \geq (F^{-1})_{ii}\,.
\end{equation}
In other words, the inverse of the FIM gives a lower bound on the covariance matrix. We calculate these estimates for the variance of the  parameters of the GW signals from the various sources in the catalogs, for each of the different scenarios under consideration (optimal and with gaps). The relative errors are then calculated for each of these sources as $\sqrt{\text{var}(\theta_i)}/\theta_i$, where $\text{var}(\theta_i)$ is the estimated variance of $\theta_i$. These results for the catalogs that incorporate the realistic ``delays'' are presented in Figs. \ref{fig:rel_err} and \ref{fig:rel_err_tc_omega}. We exclude sources that lead to GW signals with SNR less than $\rho_{th}$ = 8. 
% It should be noted that the Cramer-Rao  bound is only
% approached at high SNR. For low SNR signals originating from low mass systems, the error estimates from the FIM are therefore optimistic \cite{2008PhRvD..77d2001V, PhysRevD.57.4588}. This does, however, provide a good baseline for further studies, and allows us to keep the computational cost reasonable.

We plot the relative errors for the parameters $M_c$, $\eta = q/(1+q)^2$, $\chi_s = \frac{m1\chi_1 + m_2\chi_2}{m_1+m_2}$ and $D_L$ in Fig.~\ref{fig:rel_err}. For each parameter, we plot the logarithm of relative error for the optimal case along the x-axis, and the different gapped cases along the y-axis. We find that scheduled gaps do not introduce any major change in the error estimates. The scatter plots for the scheduled gaps for each parameter in Fig.~\ref{fig:rel_err} trace out a straight line whose slope is $\sim$1, meaning that the differences between errors for the optimal and gapped cases are negligibly small. This is expected as the loss in SNR is also minimal. In addition, there is no appreciable difference between the 3.5 hr/week and 7 hr/2 week gap scenarios. As such, we conclude that scheduled gaps are unlikely to have any significant impact on the parameter estimation of MBHB signals, provided that the merger does not take place within a gap. Fortunately, the large duty cycle of this scenario makes this a low probability event. 

\begin{center}
\begin{table*}%[h!]
  \centering
  \caption{The average number of detected sources with $\Delta\Omega<10$ sq. deg. or/and $\Delta D_L/D_L < 0.1$, in 4 years of observation for the different catalogs under consideration.}
  \begin{tabular}{|p{0.15\textwidth}|m{0.1\textwidth}|m{0.1\textwidth}|m{0.1\textwidth}|m{0.1\textwidth}|m{0.14\textwidth}|m{0.14\textwidth}|}
    \hline 
    
    \multirow{2}{*}{\textbf{Model}}  &  \multicolumn{2}{c|}{\textbf{$\Delta\Omega<10 \text{\,sq\,deg}$}} & \multicolumn{2}{c|}{\textbf{$\Delta D_L/D_L < 0.1$}} &  
    \multicolumn{2}{c|}{\textbf{$\Delta\Omega<10 \text{\,sq\,deg\,and\,}D_L/D_L < 0.1$}}\\
    \cline{2-7}
            & \textbf{Light} & \textbf{Heavy} & \textbf{Light} & \textbf{Heavy} & \textbf{Light} & \textbf{Heavy}\\
    \hline \hline
    \multicolumn{2}{|l|}{\emph{\textbf{Optimal}}} \\ \hline
    SN-Delays & 0 & 19 & 0 & 12 & 0 & 10 \\
    % \cline{1-7}
    noSN-Delays & 7 & 7 & 8 & 4 & 2 & 4 \\
    % \cline{1-7}
    SN-short Delays & 0 & 447 & 0 & 225 & 0 & 102 \\
    % \cline{1-7}
    noSN-short Delays & 11 & 401 & 13 & 265 & 4 & 119 \\
    \cline{1-7}
    \multicolumn{2}{|l|}{\emph{\textbf{Scheduled gaps 3.5 hr/week}}} \\ \hline
    SN-Delays & 0 & 15 & 0 & 9 & 0 & 8 \\
    noSN-Delays & 6 & 6 & 7 & 3 & 2 & 3 \\
    SN-short Delays & 0 & 317 & 0 & 190 & 0 & 68 \\
    noSN-short Delays & 9 & 275 & 10 & 181 & 4 & 74 \\
    \cline{1-7}
    %  & & & & & & \\
    \multicolumn{2}{|l|}{\emph{\textbf{Scheduled gaps 7 hr/2 week}}} \\ \hline
    SN-Delays & 0 & 13 & 0 & 8 & 0 & 7 \\
    noSN-Delays & 6 & 5 & 7 & 3 & 2 & 3 \\
    SN-short Delays & 0 & 290 & 0 & 153 & 0 & 59 \\
    noSN-short Delays & 9 & 266 & 10 & 170 & 4 & 71 \\
    \cline{1-7}
    \multicolumn{2}{|l|}{\emph{\textbf{Unscheduled gaps}}} \\ \hline
    SN-Delays & 0 & 9 & 0 & 6 & 0 & 4 \\
    noSN-Delays & 2 & 3 & 3 & 2 & 1 & 2 \\
    SN-short Delays & 0 & 186 & 0 & 115 & 0 & 43 \\
    noSN-short Delays & 7 & 146 & 7 & 111 & 2 & 40 \\
    \cline{1-7}

  \end{tabular}
  \label{tab:sky_DL_error}
 \end{table*}
\end{center}

The estimated errors for $D_L$, however, are quite concerning. For a significant number of sources from the catalog, the error in $D_L$ is greater than $D_L$ itself. This can be attributed to the use of PhenomD waveform models, rather than models that incorporate higher modes and/or precession. The accuracy of distance measurements is expected to improve with the inclusion of these effects. In Fig.~\ref{fig:rel_err_tc_omega}, we show  the estimated errors for the parameters $t_c$ and sky position, $\Delta\Omega = 2\pi \left(\text{var}(\lambda)\text{var}(\beta)-\text{cov}(\lambda,\beta)^2\right)$ \cite{Cutler1998}, where $\text{cov}(\lambda,\beta)$ is the covariance of the latitude and longitude ($\beta$ and $\lambda$).  These two parameters are shown separately, as tight bounds in these parameters are a prerequisite  for efficiently performing multimessenger (gravitational and electromagnetic) observations \cite{PhysRevD.102.084056,PhysRevD.93.024003,Tamanini:2016zlh}. A sky localization of $<10$ square degrees (shown as the red line) is usually considered the minimal threshold for this purpose~\cite{Tamanini:2016zlh,Belgacem:2019pkk}. 
We compute the average number of detectable sources with $\Delta\Omega<10$ sq. deg. and/or $\Delta D_L/D_L< 0.1$, seen during the observation period, for the various astrophysical models and gap scenarios (Table \ref{tab:sky_DL_error}). When compared with Table \ref{tab:detection_rates}, it is evident that only a subset of the detected sources show good localization in sky position and luminosity distance. The heavy seed catalogs are generally better in this regard than the light seed ones.
In more detail, the light seed models predict that 
less
than a handful of sources may be detected with good sky position and distance
measurements. While expected from the low SNR characterizing
binaries in these light seed scenarios~\cite{2020ApJ...904...16B,Barausse:2020gbp},
this may bode ill for the the prospects of using MBHBs to do cosmography (see e.g. \cite{Tamanini:2016zlh,Belgacem:2019pkk}).
The situation is comparatively brighter for the heavy seed models, especially 
for short delays (as expected e.g. from the results of \cite{PhysRevD.93.024003}, see especially
models Q3d and Q3nod in Figs. 7 and 12 therein).
As for the impact of missing data, as expected, unscheduled gaps have a much greater effect than the scheduled ones. Fortunately, we will be able to observe at least a few sources with good sky localization that can be observed using by other observatories. As for $t_c$, it can be estimated with an error of a few hours for most of the sources.

To examine the effect of the different gap scenarios more closely, we compute the probability distributions for the ratio of errors, $r(\theta)$, obtained in presence/absence of gaps. We define 
\begin{equation}\label{eq:ratio_error}
    r(\theta_i) = \sqrt{\frac{\text{var}(\hat{\theta}_i)|_{\text{gap}}}{\text{var}(\hat{\theta}_i)|_{\text{opt}}}}
\end{equation}
where $\text{var}(\hat{\theta}_i)|_{\text{gap}}$ is the variance in  presence of gaps and $\text{var}(\hat{\theta}_i)|_{\text{opt}}$ is the variance without gaps for the parameter $\theta_i$. Large values of $r(\theta)$ indicate larger effects from the gaps. We plot the probability distribution of $r(\theta_i)$ in Fig.~\ref{fig:relopt_err}.

Naturally, data loss will always worsen the parameter estimation, and as a consequence $r(\theta_i)\geq1$. 
However, the cases of scheduled and unscheduled gaps differ. For the two scheduled gap scenarios, we  do not observe any significant rise in the errors.
With unscheduled gaps, there appear instead
long tails towards large values of $r(\theta)$. The magnitude of the errors will, of course, depend upon the position of the gaps relative to the signal. The main reason for this disparity between scheduled and unscheduled gaps is the difference in their respective duty cycles, as discussed before. The low duty cycle of the unscheduled gaps increases the chance of a merger taking place within a gap, and the transient nature of MBHB signals (especially for high-mass systems) ensures that a significant part of the signal, if not most, is lost in such an event. For scheduled gaps, the duty cycle is $\gtrsim 95\%$, which means that the mergers are less likely to happen within gaps. Even if this happens, the loss in SNR is unlikely to be as severe as for unscheduled gaps, due to the relatively short duration of maintenance operations (3.5 or 7 hours). It can, therefore, be concluded that unscheduled gaps are more likely to induce larger errors than scheduled gaps.
The results of the FIM calculations for the catalogs incorporating the ``short delays'' models present a similar picture. These are included in Appendix \ref{short_errors_fim}.

\begin{figure*}
\subfloat[Placement of gaps for the \textit{Heavy} system. \label{fig:PE_wvf_td_heavy}]{%
  \centering
  \includegraphics[height=5.2cm,width=.49\linewidth]{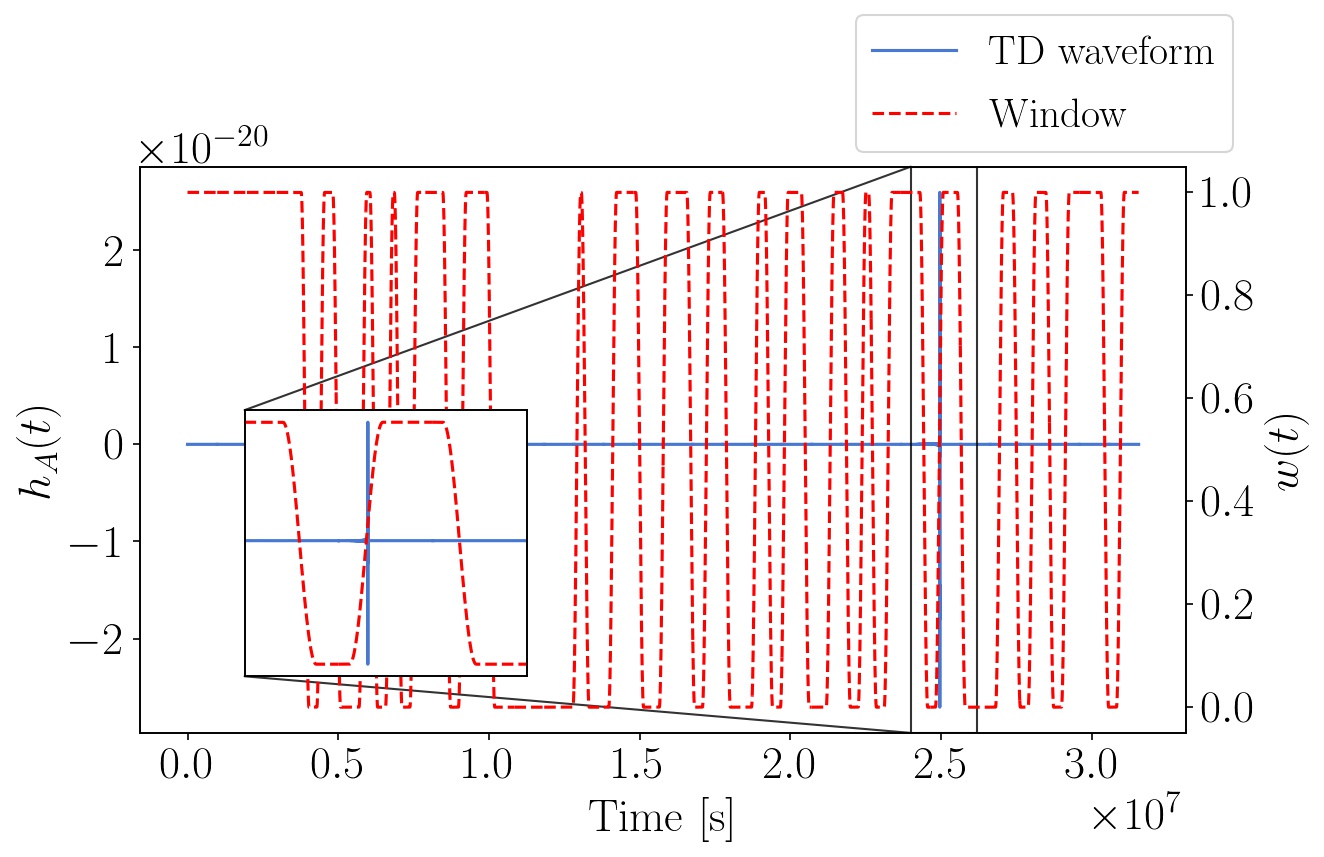}%
}\hfill
\subfloat[Frequency domain waveforms for the \textit{Heavy} system, for the optimal and gapped cases. \label{fig:PE_wvf_fd_heavy}]{%
  \centering
  \includegraphics[height=5.cm,width=.49\linewidth]{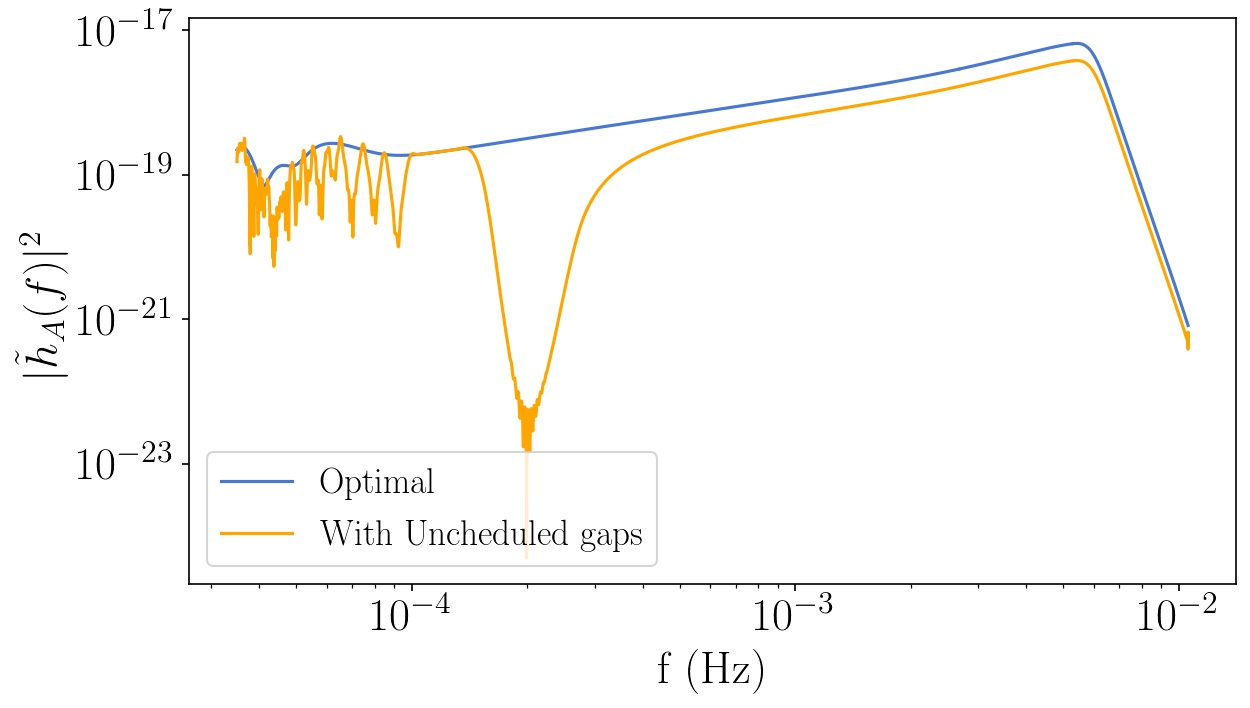}%
}\hfill
\subfloat[Placement of gaps for the \textit{Light} system. \label{fig:PE_wvf_td_light}]{%
  \centering
  \includegraphics[height=5.2cm,width=.49\linewidth]{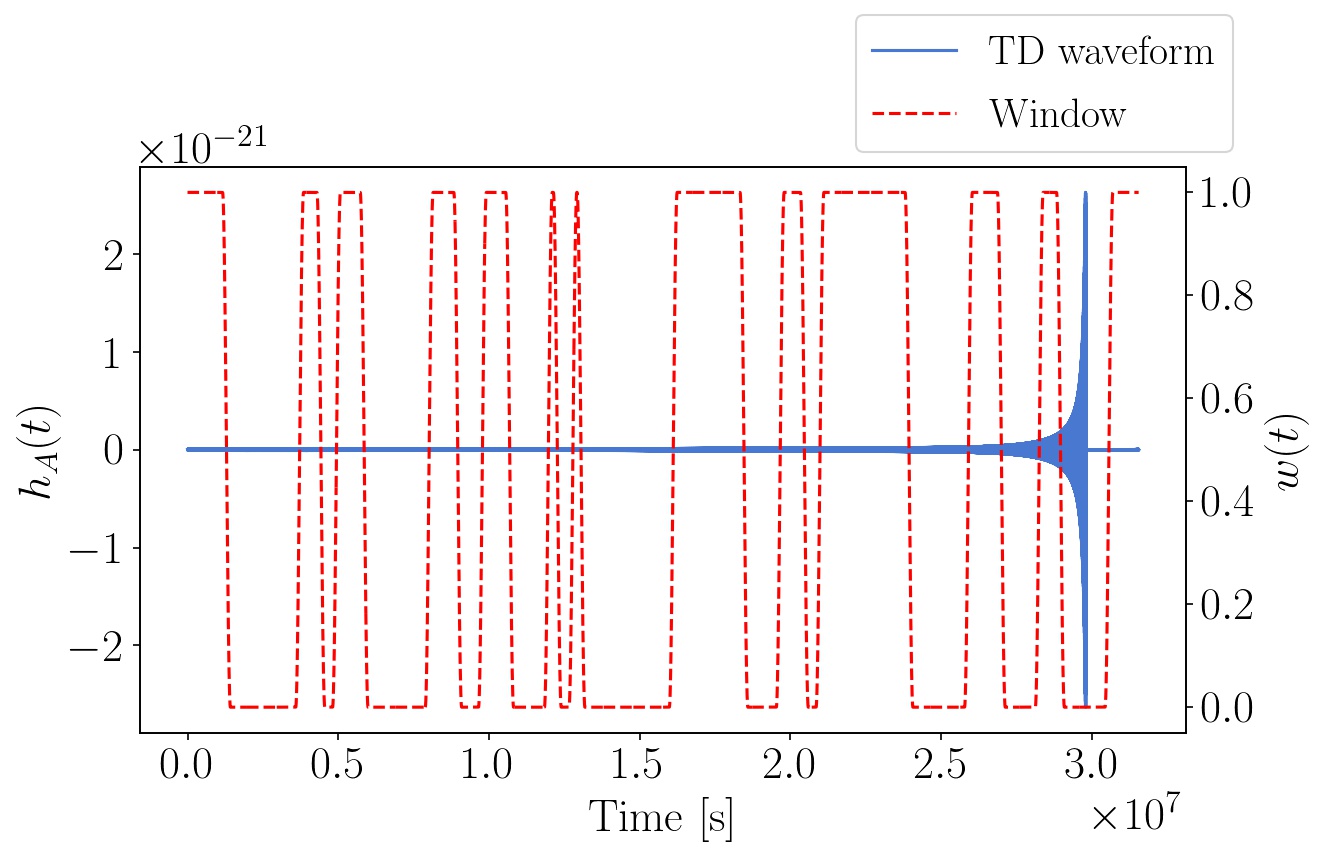}%
}\hfill
\subfloat[Frequency domain waveforms for the \textit{Light} system, for the optimal and gapped cases. \label{fig:PE_wvf_fd_light}]{%
  \centering
  \includegraphics[height=5.cm,width=.49\linewidth]{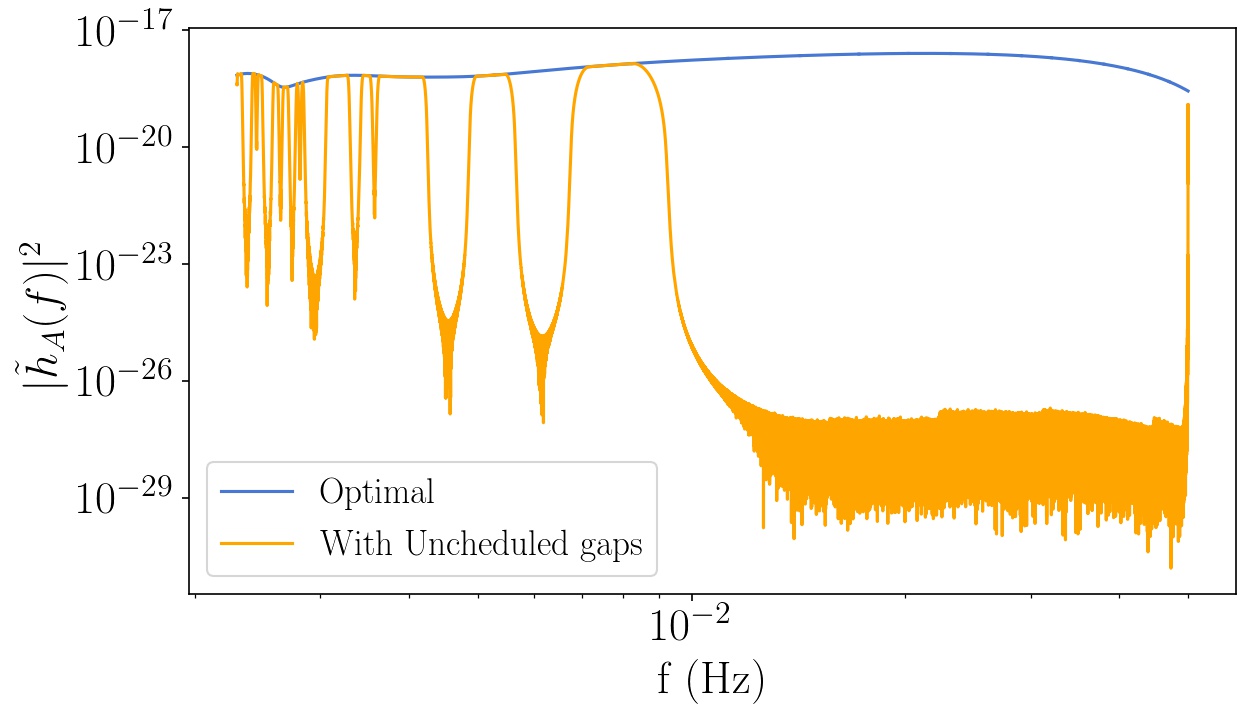}%
}
\caption{Gap placement for the \textit{Heavy} and \textit{Light} systems. The merger for the \textit{Heavy} system is shown inset in (a). The smooth transitions have effect the merger even though it lies outside the gap.}
\label{}
\end{figure*}

\section{Bayesian Parameter Estimation}\label{bayesian_PE}

 \begin{table}[ht]
  \centering
  \caption{The source parameters for the \textit{Heavy} and \textit{Light} systems.}
  \begin{center}
  \begin{tabular}{|m{0.6\linewidth}|c|c|}
    \hline 
    \centering{Parameter} & \textit{Heavy} & \textit{Light} \\ \hline
    Mass of primary, $M_1$ ($M_\odot$) & 2599137.03 & 8642.32 \\ \hline
    Mass of secondary, $M_2$ ($M_\odot$) & 1242860.68 & 587.23 \\ \hline
    Spin of primary along orbital angular momentum, $\chi_1$ & 0.75 & 0.95 \\ \hline
    Spin of secondary along orbital angular momentum, $\chi_2$ & 0.62 & 0.11 \\ \hline
    Time to coalescence, $t_c$ (seconds) & 24960000 & 29820512 \\ \hline
    Luminosity Distance, $D_L$ (Mpc) & 56005.78 & 3670.21 \\ \hline
    Inclination, $\iota$ & 1.22 & 3.06 \\ \hline
    Ecliptic longitude, $\lambda$ & 3.51 & 3.60 \\ \hline
    Ecliptic latitude, $\beta$ & 0.29 & 0.37 \\ \hline
    Polarization, $\psi$ & -0.20 & 2.11 \\ \hline
  \end{tabular}
  \label{tab:PE_heavy_light_params}
  \end{center}
 \end{table}

We choose two different systems for our Bayesian study: (i) a high mass system (chirp mass $\sim 10^6 M_\odot$) (ii) a low mass system (chirp mass $\sim 10^3 M_\odot$). These will, henceforth, be referred to as \textit{Heavy} and \textit{Light} systems respectively. The full set of parameters of these systems is shown in Table \ref{tab:PE_heavy_light_params}. 
As already mentioned, the scheduled gaps do not typically lead to  significant SNR losses, and that is also the case for both these systems. However, unscheduled gaps lead to a marked decrease in the SNR. We examine how the posterior distributions change under the effect of these unscheduled gaps. The gaps are applied as described in Sec.~\ref{data_prepare}. The signals for the optimal and gapped cases are compared in the frequency domain and are shown in Figs.~\ref{fig:PE_wvf_fd_heavy} and \ref{fig:PE_wvf_fd_light}, while the placement of gaps in the time domain is shown in Figs. \ref{fig:PE_wvf_td_heavy} and \ref{fig:PE_wvf_td_light}.

For the \textit{Heavy} system, the last gap ends half a day before  merger. The smoothing at the ends of the gap (see Sec.~\ref{data_prepare}) affects the merger, as well as part of the ringdown (Fig.~\ref{fig:PE_wvf_td_heavy}). The frequency domain TDI-A amplitudes of the optimal and gapped cases are compared in Fig.~\ref{fig:PE_wvf_fd_heavy}. The dips correspond to the gaps in the time domain. The effect of windowing is clearly observed in the high frequency region, as the gapped waveform has lower power than the optimal case. The SNR for the optimal case is $308.8$ which decreases to $178.5$ under the impact of gaps. The SNR accumulates very rapidly around the merger for this source, and positioning the gap closer to the merger will result in a very rapid decrease of its value. In fact, if the gap placement is such that the merger takes place within an unscheduled gap, then the chances of observing the source are very slim, as the SNR is likely to be less than $\rho_{th}$. This is in stark contrast with lower mass systems, as we shall see.

\begin{figure*}
        \centering
        \includegraphics[width=\textwidth]{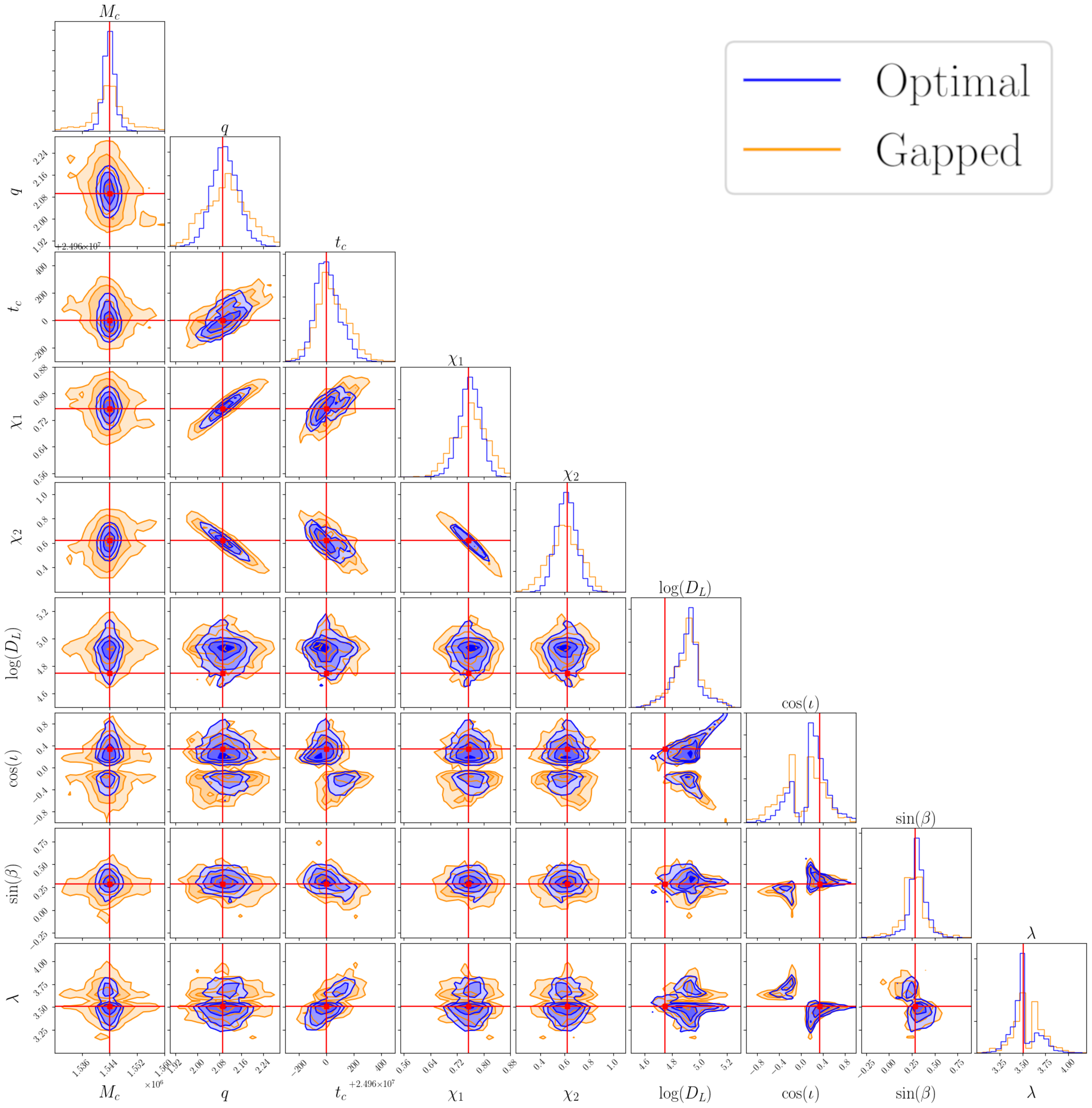}
        \caption{Posterior distribution for the \textit{Heavy} system in Table \ref{tab:PE_heavy_light_params}. The posteriors are shown in blue for the optimal case, and in orange in  presence of unscheduled gaps. The solid red lines represent the injected values.}
        \label{fig:heavy_PE}
\end{figure*}

\begin{figure*}
        \centering
        \includegraphics[width=\textwidth]{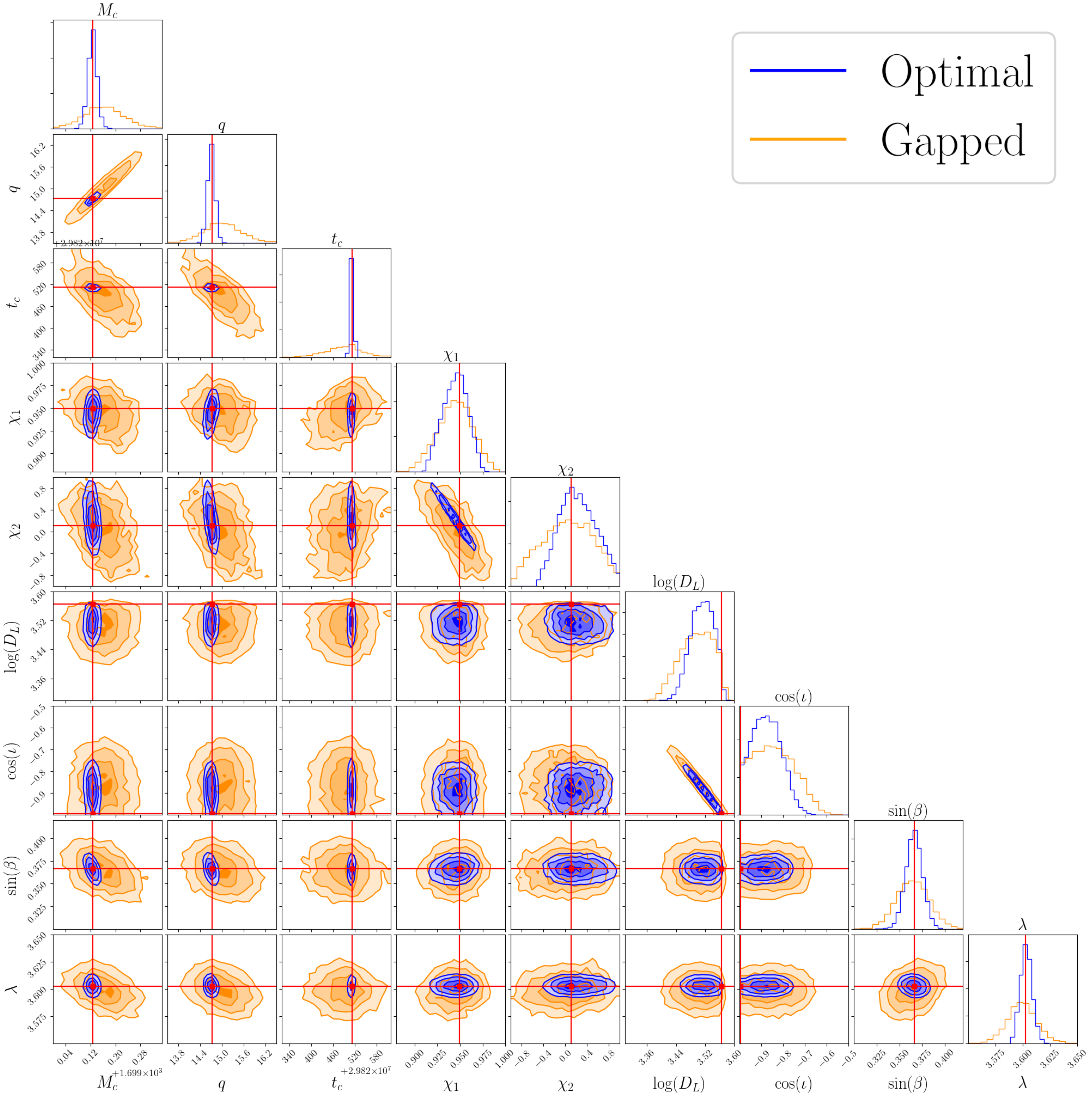}
        \caption{Posterior distributions for the \textit{Light} system of Table \ref{tab:PE_heavy_light_params}. The posteriors are shown in blue for the optimal case, and in orange in  presence of unscheduled gaps. The solid red lines represent the injected values.}
        \label{fig:light_PE}
\end{figure*}

For the \textit{Light} system, the merger takes place within a gap (Fig.~\ref{fig:PE_wvf_td_light}). A large portion of the high frequency part of the waveform is inaccessible due to the gaps. The SNR decreases from $81$ in the optimal case to $45$ in the presence of unscheduled gaps and is distributed more evenly across the LISA band, unlike for the \textit{Heavy} system. 

Note that we do not sample the space spanned by the parameters in Table \ref{tab:PE_heavy_light_params}, but rather carry out our search with the parameters ($M_c, q, \chi_1, \chi_2, t_c, \log D_L, \cos \iota, \sin \beta, \lambda, \psi$) (see Sec.~\ref{PE_des}).
We choose uninformative (i.e. uniform) priors on these new parameters. The range of these priors is such that the true value of the parameters of the GW signal are far from the boundary values (see Table \ref{tab:PE_prior} for details). 

The sampling was performed using the package \texttt{PTMCMCSampler} (see Sec.~\ref{PE_des}).
\begin{table}[h]
\begin{tabular}{|l|c|c|}
\hline
\multirow{ 2}{*}{Parameter} & \multicolumn{2}{c|}{Prior}   \\ 
                             & \multicolumn{1}{c}{Heavy System} & Light System \\ \hline\hline
$M_c~\left[M_\odot\right]$ &  $\mathcal{U}\left[10^6, \,2\times10^6\right]$ &    $\mathcal{U}\left[1.5\times10^3, \,2\times10^3\right]$\\ 
$q$                          &  $\mathcal{U}\left[1, \,3\right]$ &  $\mathcal{U}\left[10, \,16\right]$  \\ 
$\chi_1$, \,$\chi_2$        & $\mathcal{U}\left[-1, \,1\right]$  &  $\mathcal{U}\left[-1, \,1\right]$  \\ 
$t_c~\left[\mathrm{s}\times10^7\right]$ &  $\mathcal{U}\left[2.476,\, 2.516\right]$ &  $\mathcal{U}\left[2.962, \,3.002\right]$  \\ 
$\log D_L$                    &  $\mathcal{U}\left[4.5, \,5.5\right]$ & $\mathcal{U}\left[3.2, \,4\right]$   \\ 
$\cos\iota$                          & $\mathcal{U}\left[-1,\,1\right]$  &  $\mathcal{U}\left[-1,\,1\right]$  \\ 
$\sin\beta$                          &  $\mathcal{U}\left[-1,\,1\right]$ &  $\mathcal{U}\left[-1,\,1\right]$  \\ 
$\lambda$                          &  $\mathcal{U}\left[0,\,2\pi\right]$ &  [$\mathcal{U}\left[0,\,2\pi\right]$  \\
$\psi$                          & $\mathcal{U}\left[0,\,\pi\right]$  &  $\mathcal{U}\left[0,\,\pi\right]$ \\ \hline
\end{tabular}
\caption{Summary of the prior densities used for the parameter estimation of both the heavy and the light systems under study.}
\label{tab:PE_prior}
\end{table}

For this study, we used 56 parallel temperature chains, with temperature difference between successive chains following a geometric spacing. This was found to be sufficient for our study. We used a proposal distribution consisting of weighted sub-proposals, including Adaptive Metropolis \cite{10.2307/3318737}, Single Component Adaptive Metropolis \cite{Haario2005SCAM} and Differential Evolution jumps \cite{Braak2008}. Each of the chains was run for 2.5 to 3 million steps. The effective number of samples, however, is less than the total number of steps, as one has to exclude the burn-in region and take into account the auto-correlation length. We found that in all cases we retain an effective sample size of $\gtrsim 2500$ after excluding the burn-in samples and down sampling by the auto-correlation length.

Fig.~\ref{fig:heavy_PE} and \ref{fig:light_PE} show a comparison between the posterior distributions for the optimal and gapped cases, for the \textit{Heavy} and \textit{Light} systems respectively. The results for the \textit{Heavy} system show degeneracies among the extrinsic parameters. This is because of the short signal duration (less than one year). The effect of the extrinsic parameters is imprinted on the waveform through the LISA response function, which depends on both time and frequency (see Sec.~\ref{reponse_and_noise}). The spacecraft are effectively stationary during the lifetime of the signal, and thus the time dependence of the response does not play a prominent role. This leads to the secondary maxima in the sky position parameters $\sin\beta$ and $\lambda$ (see \cite{2020arXiv200300357M} for details). The \textit{Light} system, instead, remains in the LISA band for longer (Fig.~\ref{fig:PE_wvf_td_light}). The time dependence of the LISA response is therefore in full effect and breaks the degeneracies in the sky position. Because of this, we only see the primary maxima in the sky position.
For similar reasons, the posterior for the inclination is multimodal for the \textit{Heavy} system. Recall that we restrict ourselves to the dominant $(2,\pm2)$, and the inclusion of higher harmonics should help break (at least partially) these degeneracies \cite{PhysRevD.81.064014, 2008PhRvD..77b4030T, ArunIyer2007, 2008PhRvD..78f4005P}.

The gaps result in broader posteriors for most of the sampled parameters. In fact, $r(\theta)$ lies in the range $\sim$ 1.1 -- 3 for the \textit{Heavy} system and $\sim 1.3$ -- 12 for the \textit{Light} system. The effect of gaps is larger for the \textit{Light} system, because data from the late inspiral and merger phases, which contribute heavily to the determination of the mass ratio, spins and merger time, is lost. The peaks of the recovered posteriors, however, are close to the true values of the parameters for both systems. The time of coalescence of the binary, $t_c$, is recovered very well for the \textit{Heavy} system. Even for the \textit{Light} system, where the merger takes place within a gap, $t_c$ is estimated to be within a period of 3 minutes. We were also able to recover the sky position quite well in the presence of gaps, for both  systems. Though the high value of $r(\theta)$ for the \textit{Light} system is a cause of concern, it is still possible to estimate the source parameters. The same cannot be said for the \textit{Heavy} source, if the merger were to take place within a gap, as the SNR would fall below the detection threshold $\rho_{th}$. 

In summary, the results of the full Bayesian parameter estimation for the two systems paint a somewhat promising picture. For the transient signal from the \textit{Heavy} source, the posterior distribution can be recovered even if the the gap is extremely close to the merger, as long as it does not completely encompass it. For the long signal from the \textit{Light} system, the posterior distributions can be recovered even if the merger takes place completely inside a gap. The peak values for most of the source parameters (except $\log D_L$ and $\cos \iota$) lie quite close to the true values. As for the prospects of multi-messenger astronomy, we conclude that  MBHB signals that have a sufficient SNR in the inspiral phase can be tracked for a long time, and electromagnetic observatories can be alerted prior to possible merger events. The fact that $t_c$ and the sky position can be recovered reasonably well (e.g. to within a range of few minutes for $t_c$) for these sources bodes well for future joint operations between LISA and electromagnetic telescopes. However, as we saw in Sec.~\ref{errors_fim}, localization of a source in the sky may be quite difficult depending on the gap distribution and the SNR of the signal. Although it is unlikely that we would be able to recover good accuracy in sky position for all of the sources, we should be able to do it for some of them. Tracking accurately the signal (and $t_c)$  also allows for scheduling `protected periods'
where maintenance operations are put on hold, so as to avoid jeopardizing direct GW observations 
of the merger.

\section{Conclusions}\label{conclusion}

In this work, we have investigated the impact of data gaps in the LISA data stream on the detectability and parameter estimation of MBHB signals. We performed our analysis exploring both the parameter space of the GW signals generated by MBHBs in the presence and absence of gaps. Considering the data gaps, we have focused on two cases based on the experience of the LISA Pathfinder mission. For the first case, which simulates the data 
loss due to scheduled space-craft maintenance, we have assumed data gaps of 3.5 hours every week or 7 hours every two weeks. The second case 
considers instead longer unscheduled data gaps (of three days each, repeated 
so that the overall duty cycles is 75 \%), which may result from
 unforeseen anomalies and/or failures in one or more components in the instrument.
For the population of MBHBs,
 we have  considered eight different astrophysical scenarios bracketing modeling uncertainties. Of these, four are ``heavy'' seed models with MBHB masses in the range $\sim 10^5 - 10^8 M_\odot$, and four are ``light'' seed models with MBHB masses in the range $\sim 10^3-10^5 M_\odot$. 

Our analysis is based on windowing the data in the time domain with a smooth window function. While this methodology is quite straightforward, it might not produce optimal results compared to other data gap treatments. The reason is that the choice of the window function matters for each individual analysis scenario, and more importantly, more useful data are being lost due to the window tapering. Nevertheless, our work here does not focus on studying the optimal gap treatment, but rather on making a baseline assessment of the impact of data gaps on the detectability of MBHB signals with LISA. Alternative methods for gap treatment can be found in~\cite{PhysRevD.100.022003} and in~\cite{Blelly2021oim}. We then assess the impact of data gaps of different lengths by adopting different measures, such as the SNR loss. At the same time, we estimate the parameter estimation degradation by computing the FIM. It is worth noting that this work focuses on detectability based on a given SNR threshold. We assume that there are no further complications, and restrict ourselves to individual events with no overlapping signals. The effect of gaps on actual detection pipelines will be investigated in future studies.

We have found that the impact of gaps on MBHB signals is greater as the gap gets closer to the merger. This was observed for both heavy and light seed binaries. For the former, the SNR of the signal may be too low for detection, if the merger takes place inside a gap. For  lighter systems, although the optimal SNR is typical low, it is spread over the observation duration and can be detected even if the merger happens within a gap. It is, however, worth noting that for  low mass systems, the low optimal SNRs will lead to low number of detections, even in the absence of gaps.

We have found that the scheduled maintenance gaps are unlikely to have a significant impact on detection and parameter estimation, provided that they do not encompass the merger, (something that could be arranged for during operations by defining protected periods). Moreover the specific choice of scenario for the scheduled gaps, i.e. 3.5 hours per week or 7 hours per 2 weeks, does not affect the detectability of the signal appreciably (see table~\ref{tab:detection_rates} and~\ref{tab:sky_DL_error}). The 7 hours gap per 2 weeks scenario is marginally better, as less data is lost due to the smoothing at the ends of the gaps. The unscheduled gaps have instead a much greater effect, owing to their potentially longer duration, which increases the chance of the merger happening within one of them. We have reached this conclusion by performing FIM calculations, which show that the errors on the source parameters remain quite small even with scheduled gaps, while they can become significant for unscheduled gaps. 

Finally, for the sake of completeness, we have performed a full Bayesian analysis for two representative cases: a heavy and a light MBHB system. We have sampled the posterior distribution for the optimal case (without any data gaps), and a case where the data are interrupted by unscheduled gaps. The motivation for this investigation was to test for secondary maxima on the posterior hypersurface induced by the data interruptions. For the chosen gap realizations, the results are promising. We have found that the presence of gaps results in a broadening of the posterior distributions. Other realizations may however deteriorate further the results, particularly for the heavy system, for which a merger within a gap would make detection impossible or very challenging.

Our observations lead us to conclude that MBHB signal analysis will be impacted the most by the unscheduled gaps in the data stream. Fortunately, estimation of the source parameters is still possible for such signals, at least as long as gaps happen to fall far from the merger. 
Mergers happening within unforeseen data gaps may however lead to missed detection of events, especially for MBHB sources with masses at the higher part of the spectrum.

\acknowledgements
This work is implemented as a contribution to {\it Artefacts group} within Work Package 2 (WP2) of LISA Science Group (LSG). K. D. greatly acknowledges the support for high-performance computing time at the Padmanabha cluster, IISER Thiruvananthapuram, India.
E. B.  acknowledges financial support provided under the European Union's H2020 ERC Consolidator Grant ``GRavity from Astrophysical to Microscopic Scales'' grant agreement no. GRAMS-815673.
N. Korsakova acknowledges the support from CNES fellowship and the support by the LABEX Cluster of Excellence FIRST-TF (ANR-10-LABX-48-01), within the Program "Investissements d'Avenir” operated by the French National Research Agency (ANR).

\appendix

\section{Noise model}\label{app:noise}

\begin{figure}
    \centering
    \includegraphics[width=0.45\textwidth]{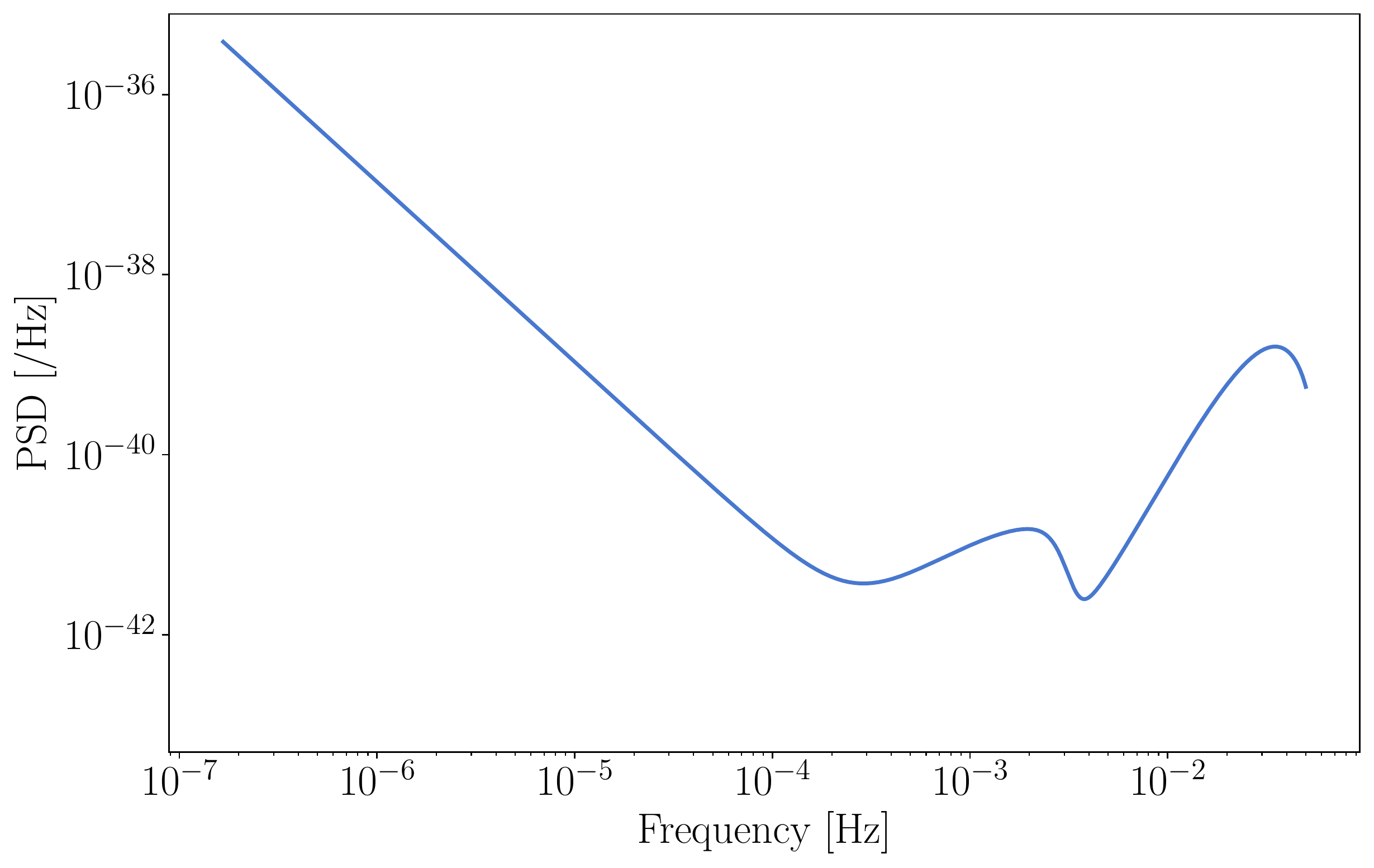}
    \caption{PSD of \texttt{SciRDv1} accounting for GB confusion}
    \label{fig:scirdv1_gb}
\end{figure}

\begin{figure}[htb!]
    \centering
    \includegraphics[width=0.45\textwidth]{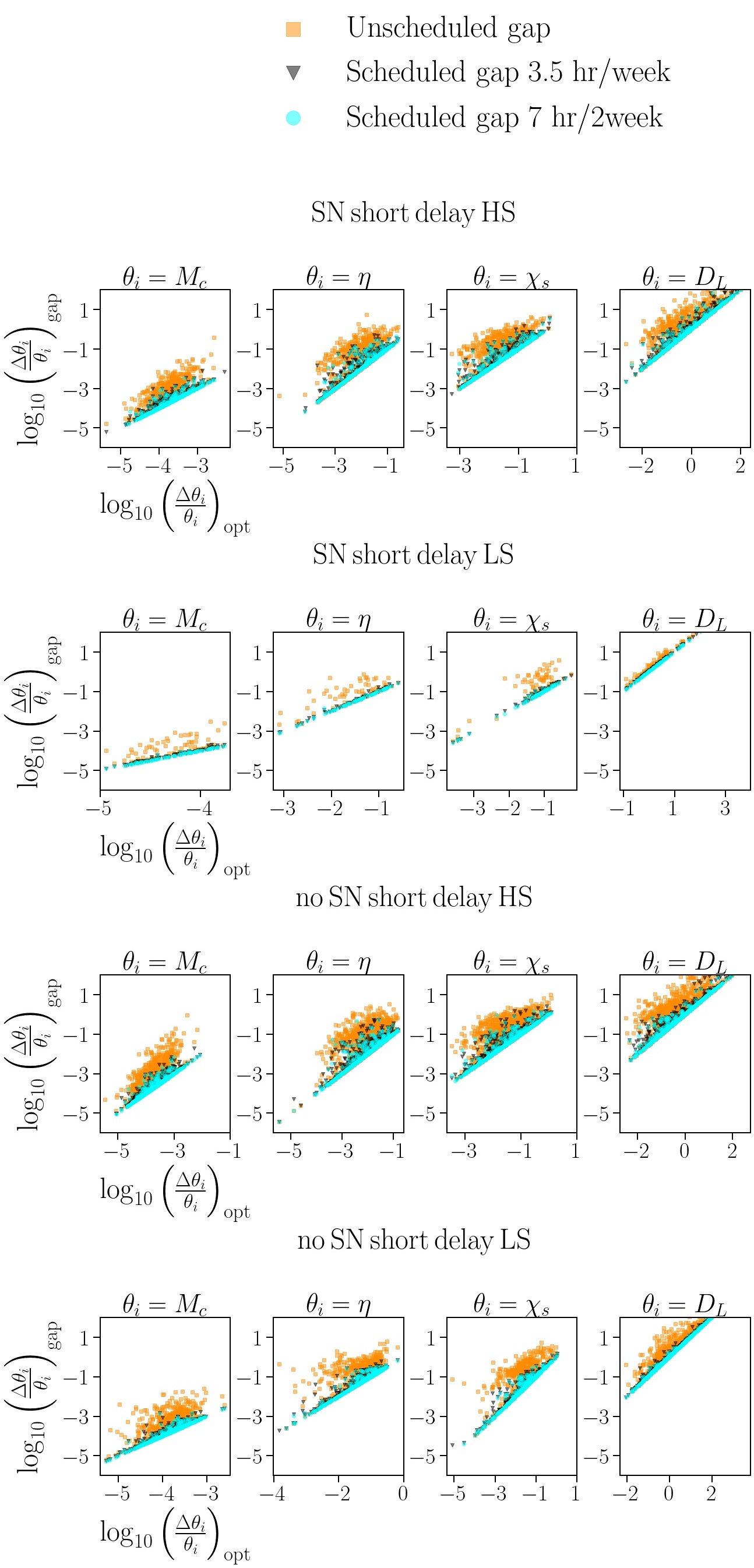}
    \caption{Comparison of relative errors for the optimal and gapped cases for the sources in the `short delays' catalogs from Table \ref{tab:detection_rates}. Along the x-axis, we plot the logarithm of relative errors for the optimal case while along y-axis, we plot the same for the different gap scenarios. 
    }
    \label{fig:rel_err_short}
\end{figure}

The noise model used in this study (\texttt{SciRDv1}) is described in \cite{LISA_SciRDv1}. We are interested in the noise PSD for the three TDI channels.
\begin{subequations}
\begin{eqnarray}
    S_n^A(f) =&& 8 \sin^2 \left(\frac{f}{f_*}\right)\Bigg[2S_{\text{pm}}\left(3 + \cos\left(\frac{2f}{f_*} \right) + \cos\left( 2\frac{f}{f_*}\right)\right)\nonumber\\ 
    &&+ S_{\text{op}}\left(2 + \cos\left(\frac{f}{f_*} \right) \right) \Bigg] 
\end{eqnarray}
\begin{equation}
    S_n^A(f) = S_n^E(f)
\end{equation}
\begin{eqnarray}
    S_n^T(f) =&& 16 \times S_{\text{op}} \sin^2\left(\frac{f}{f_*} \right)\left(1 - \cos\left(\frac{f}{f_*} \right) \right)\nonumber \\ 
    &&+ 128 \times S_{\text{pm}} \sin^2\left(\frac{f}{f_*} \right) \sin^4\left(\frac{f}{2f_*} \right)
\end{eqnarray}
\end{subequations}
where $S_n^{A,E,T}(f)$ are the noise PSDs of the TDI A, E, T channels and $S_{\text{pm}}(f)$ and $S_{\text{op}}(f)$ are respectively the test-mass acceleration noise PSD (\ref{eq:acc_psd}) and the optical metrology system noise PSD (\ref{eq:oms_psd}),  in relative frequency units, with $f_* = c/2\pi fL$.

\begin{subequations}\label{eq:acc_oms_psd}
\begin{eqnarray}\label{eq:acc_psd}
    S_{\text{pm}}(f) =&& (3\cdot10^{-15})^2 \left(1 + \left(\frac{0.4\cdot10^{-3}}{f}\right)^2 \right)\nonumber \\ &&\cdot\left(1 + \frac{f}{8\cdot10^{-3}}\right)^4\left( \frac{1}{2\pi f}\right)^4\left(\frac{2\pi f}{c}\right)^2
\end{eqnarray}
\begin{eqnarray}\label{eq:oms_psd}
    S_{\text{op}}(f) =&& (15\cdot10^{-12})^2 \nonumber \\ && \cdot\left(1 + \left(\frac{2\cdot10^{-3}}{f}\right)^4 \right)\left(\frac{2\pi f}{c}\right)^2
\end{eqnarray}
\end{subequations}
The unresolved galactic white dwarf binary sources present in the LISA will give rise to a background noise in the frequency range 0.1 - 5 mHz \cite{Bender_1997, 1990ApJ...360...75H}. This leads to a slight decrease in LISA sensitivity. For this study, we add  this effect in the analytic noise model. The resulting \texttt{SciRDv1} noise model is shown in Fig.~\ref{fig:scirdv1_gb}.

\section{FIM results for ``short delays'' catalogs}\label{short_errors_fim}

Figures \ref{fig:rel_err_short} and \ref{fig:rel_err_tc_omega_short} shows the estimated errors for the catalogs incorporating the ``short delays'' models. In Fig.~\ref{fig:rel_err_short}, we plot the logarithm of relative errors for the optimal case along the x-axis and the gapped cases along the y-axis, for the parameters $M_c$, $\eta$, $\chi_s$ and $D_L$. In Fig.~\ref{fig:rel_err_tc_omega_short}, we plot the errors on $t_c$ and sky position $\Omega$. The results are similar to those seen in Fig.~\ref{fig:rel_err} and \ref{fig:rel_err_tc_omega}. Fig.~\ref{fig:relopt_err_short} shows the probability distribution of $r(\theta)$ (see \eqref{eq:ratio_error}). The results are similar to Fig.~\ref{fig:relopt_err}. See Sec.~\ref{errors_fim} for a detailed discussion.

\section{Estimation of the length and frequency of unscheduled gaps based on LPF data}\label{app:dfacs}

\begin{figure*}
    \centering
    \includegraphics[width=\textwidth]{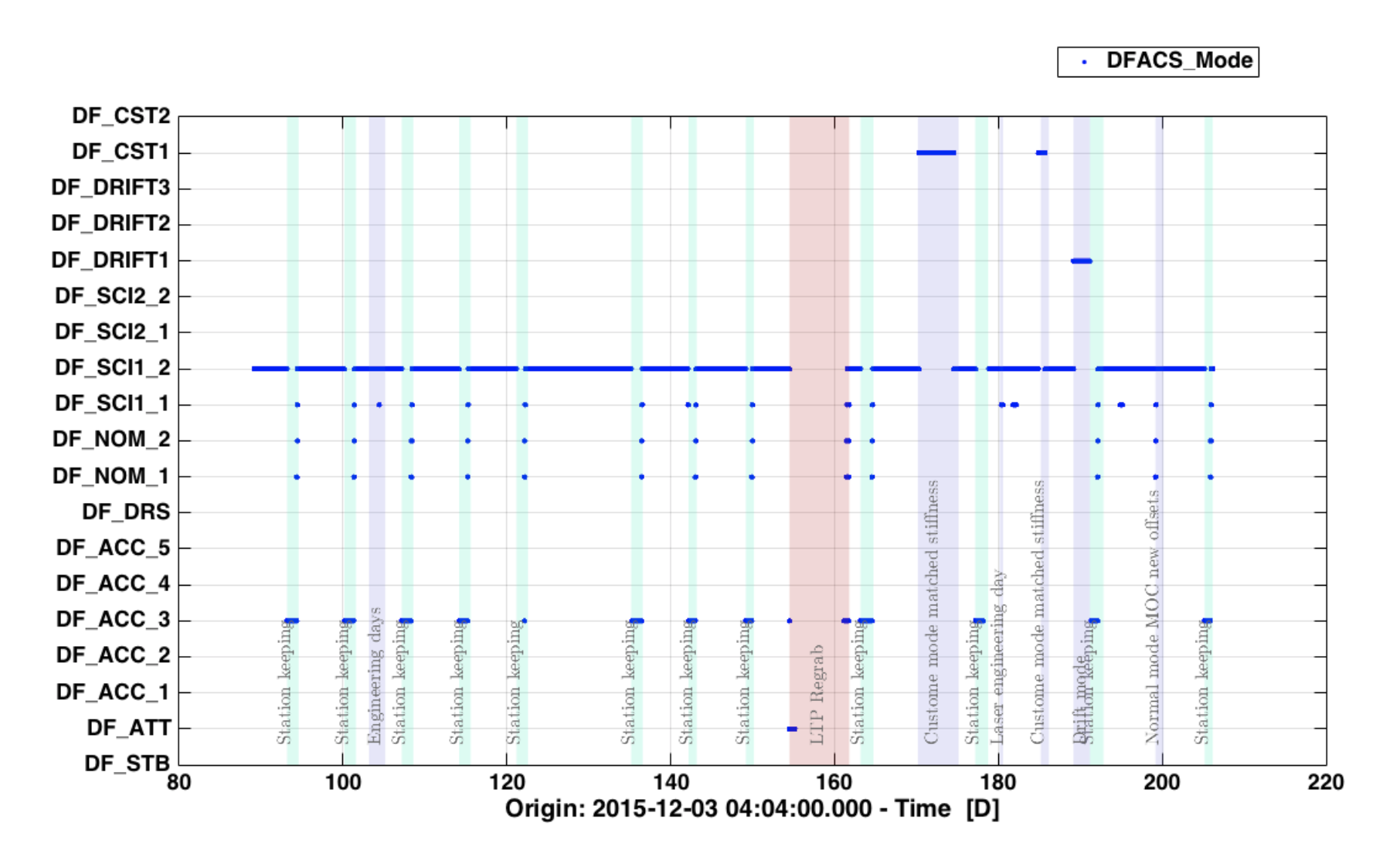}
    \caption{Summary of the activities performed during the LPF
nominal mission, which were used to estimate  the length and frequency of the unscheduled gaps. The $x$-axis indicates the number of days from the beginning of the mission. The $y$-axis indicates the DFACS mode in which the spacecraft was during the operations. The blue dot for each day defines in which mode the spacecraft has been. The green vertical lines correspond to {\it Station keeping}, the purple lines correspond to the custom modes, in which dedicated experiments have been performed. The red line indicated the days when the spacecraft was not in one of the valid modes.} 
    \label{fig:unsheduled_gap_dfacs}
\end{figure*}

\begin{figure}[hbt!]
    \centering
    \includegraphics[width=0.4\textwidth]{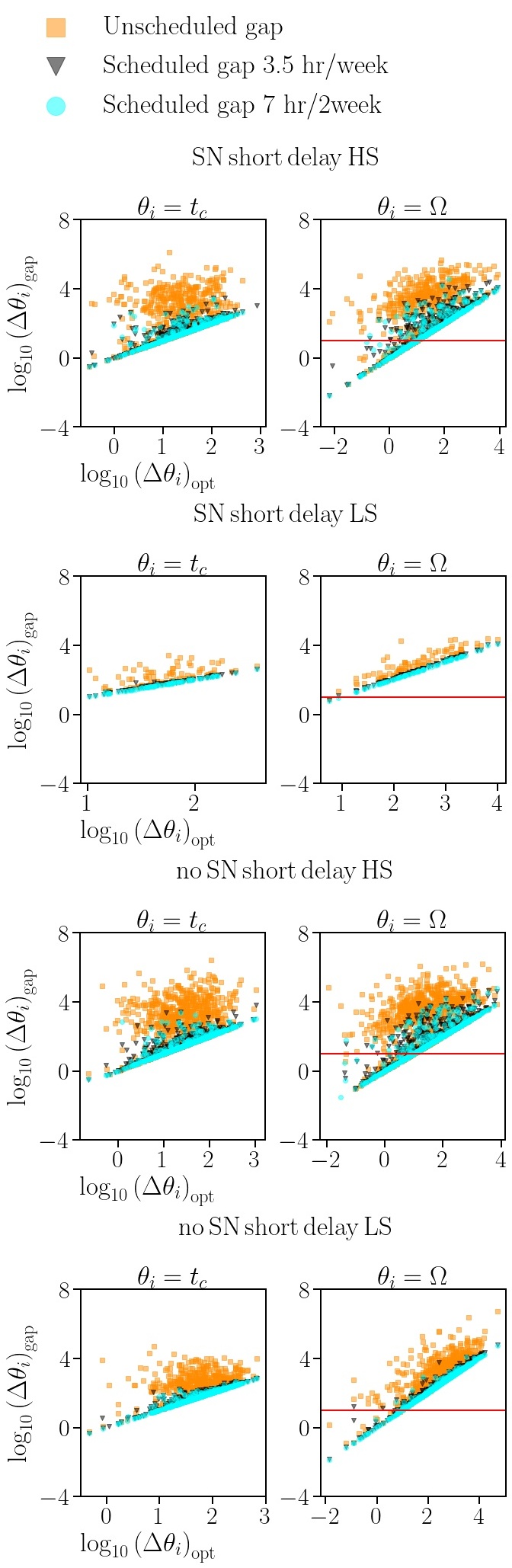}
    \caption{Same as Fig.~\ref{fig:rel_err_short} but for the merger time and sky localization.}
    \label{fig:rel_err_tc_omega_short}
\end{figure}

\begin{figure*}
    \centering
    \includegraphics[width=\textwidth]{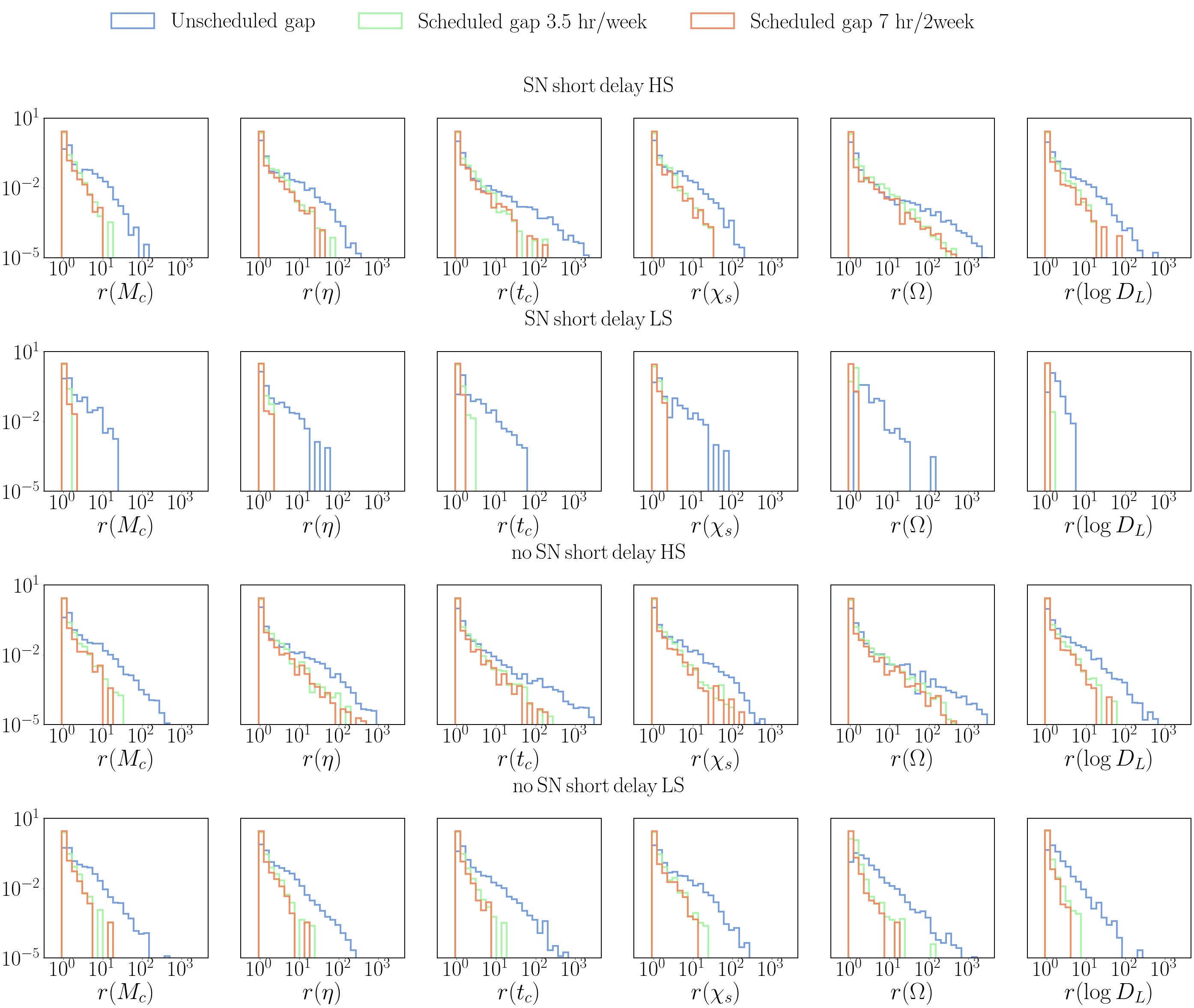}
    \caption{Probability distribution of $r(\theta)$ for the sources in `short delays' catalogs, from Table \ref{tab:detection_rates}}
    \label{fig:relopt_err_short}
\end{figure*}

The Failure Detection, Isolation and Recovery (FDIR) procedure is activated when a certain security threshold on the spacecraft is exceeded and the spacecraft goes into a safe mode. We estimate the rate and length of unplanned gaps by looking at the active modes of DFACS (Drag Free and Attitude Control System) in which LPF has been operating during the lifetime of the mission. The list of available DFACS modes is provided in Table~\ref{DFACS_modes}. Fig.~\ref{fig:unsheduled_gap_dfacs} shows the summary of the activities for the nominal mission. For each day a blue dot indicates in which mode was the spacecraft. Science operations are performed in  \verb+SCI1.2+ mode. Station keeping is performed in \verb+ACC3+ mode and is highlighted in  Figure~\ref{fig:unsheduled_gap_dfacs} with the green colour. The other planned activities  which were reflected in the mission agenda are highlighted in purple and include, for example:
\begin{itemize}
\item
drift mode free-fall experiment in \verb+DRIFT1+ mode;
\item
noise measurement and experiments in the mode similar to \verb+SCI1.2+ but with the matched stiffness which is programmed in mode called \verb+CST1+;
\item
number of engineering activities which were done in \verb+SCI1.1+ or \verb+NOM+ modes.
    \end{itemize}

We can see that the only episode when the spacecraft was not functioning as expected was when the test mass regrab happened. It is highlighted in  Figure~\ref{fig:unsheduled_gap_dfacs} with a red colour. We estimate the time when the spacecraft was not functional by measuring the time of the test mass regrab versus the time when the spacecraft was fully operational.

\begin{table}[!h]
\begin{tabular}{|r|l|l|}
\hline
0 & \verb+DF_STB+ & DFACS  Standby Mode \\
1 & \verb+DF_ATT+ & DFACS  Attitude Mode \\
2 & \verb+DF_ACC_1+ & DFACS  Accelerometer Mode 1 \\
3 & \verb+DF_ACC_2+ & DFACS  Accelerometer Mode 2 \\
4 & \verb+DF_ACC_3+ & DFACS  Accelerometer Mode 3 \\
5 & \verb+DF_ACC_4+ & DFACS  Accelerometer Mode 4 \\
6 & \verb+DF_ACC_5+ & DFACS  Accelerometer Mode 5 \\
7 & \verb+DF_DRS+ & DFACS DRS Mode \\
8 & \verb+DF_NOM_1+ & DFACS Normal Mode 1 \\
9 & \verb+DF_NOM_2+ & DFACS Normal Mode 2 \\
10 & \verb+DF_SCI1_1+ & DFACS Science Mode 1.1 \\
11 & \verb+DF_SCI1_2+ & DFACS Science Mode 1.2 \\
12 & \verb+DF_SCI2_1+ & DFACS Science Mode 2.1 \\
13 & \verb+DF_SCI2_2+ & DFACS Science Mode 2.2 \\
14 & \verb+DF_DRIFT1+ & DFACS Drift Mode 1 \\
15 & \verb+DF_DRIFT2+ & DFACS Drift Mode 2 \\
16 & \verb+DF_DRIFT3+ & DFACS Drift Mode 3 \\
17 & \verb+DF_CST1+ & DFACS Custom Mode 1 \\
18 & \verb+DF_CST2+ & DFACS Custom Mode 2 \\
\hline
\end{tabular}
\caption{Available DFACS modes\label{DFACS_modes}}
\end{table}

\FloatBarrier
\end{document}